\documentclass[12pt]{iopart}

\usepackage{epsfig,amssymb, graphicx}
\usepackage{deluxetable}

\def\msol{M_\odot}
\def\msun{M_\odot}
\def\mjup{M_{\rm Jup}}
\def\rjup{R_{\rm Jup}}
\def\finc{F_{\rm inc}}
\def\mearth{\,{\rm M}_\oplus}
\def\rearth{\,{\rm R}_\oplus}
\def\mnep{\,{\rm M}_{\rm Nep}}
\def\mcore{\,{\rm M}_{\rm core}}

\def\simgr{\,\hbox{\hbox{$ > $}\kern -0.8em \lower 1.0ex\hbox{$\sim$}}\,}
\def\simle{\,\hbox{\hbox{$ < $}\kern -0.8em \lower 1.0ex\hbox{$\sim$}}\,}

\def\gc2{\,\,\,{\rm g}\,{\rm cm}^{-2}}             

\def\mnras{{\it MNRAS}}
\def\nat{{\it Nature}}
\def\aap{{\it Astron. Astroph.}}
\def\apj{{\it Astroph. Journal}}
\def\apjl{{\it Astroph. Journal Let.}}
\def\apjs{{\it Astroph. Journal Suppl.}}
\def\araa{{\it Annu. Rev. Astron. Astrophys.}}
\def\aj{{\it Astron. Journal}}
\def\icarus{{\it Icarus}}
\def\mnras{{\it Mon. Not. R. Astron. Soc.}}
\def\prl{{\it Phys. Rev. Lett.}}
\def\prb{{\it Phys. Rev. B}}
\def\jpf{{\it Journ. Phys. F}}
\def\jpa{{\it Journ. Phys. A}}

\def\pasp{{\it Pub. Astr. Soc. Pacific}}

\begin{document}
\bibliographystyle{vancouver}

\title{The physical properties of extrasolar planets}

\author{I Baraffe$^{1, 2}$, G Chabrier$^1$ and T Barman$^3$}

\address{$^1$ \'{E}cole normale sup\'erieure de Lyon, CRAL (CNRS), 46 all\'ee d'Italie, 69007 Lyon,\\ Universit\'e de Lyon, France}
\address{$^2$School of Physics, University of Exeter, Stocker Rd, Exeter EX4 4QL} 
\address{$^3$ Lowell observatory, 1400 West Mars Hill Road, Flagstaff, AZ 86001, USA}
\eads{\mailto{isabelle.baraffe@ens-lyon.fr}, \mailto{gilles.chabrier@ens-lyon.fr}, \mailto{barman@lowell.edu}}

\begin{abstract}
Tremendous progress in the science of extrasolar planets has been achieved since
the discovery of a Jupiter orbiting the nearby Sun-like star 51 Pegasi in 1995.
Theoretical models have now reached enough maturity to predict the
characteristic properties of these new worlds,
mass, radius, atmospheric signatures, and can be confronted with available
observations. We review our
current knowledge of the physical properties of exoplanets, internal structure
and composition,
atmospheric signatures, including expected biosignatures for exo-Earth planets,
evolution, and the impact of tidal interaction and stellar irradiation on these properties for the
short-period planets. We
discuss the most recent theoretical achievements in the field and the still
pending questions.
We critically analyse the different solutions suggested to explain 
abnormally large radii of a significant fraction of transiting exoplanets.
Special attention is devoted to the  recently discovered transiting objects in
the overlapping mass range between massive planets and low-mass brown dwarfs,
stressing the ambiguous nature
of these bodies, and we discuss the possible observable diagnostics to identify
these two distinct populations. We also review our present understanding of
planet formation and critically examine the different suggested formation
mechanisms. We expect the present review to provide the basic theoretical
background to capture the essential of the physics of exoplanet formation,
structure and evolution, and the related observable signatures.
\end{abstract}

\maketitle

\newpage

\noindent{\bf Contents} \\
\medskip
\noindent{\bf 1. Introduction}\dotfill\quad\ 3 \\
{\bf 2. A brief overview of observations}\dotfill\quad\ 4 \\
\indent{\textit{2.1. Lessons from our Solar System}} \dotfill\quad\ 4 \\
\indent{\textit{2.2. Observed properties of exoplanets}} \dotfill\quad\ 6 \\
{\bf 3. Planet formation}\dotfill\quad\ 9 \\
\indent{\textit{3.1. Core-accretion model}} \dotfill\quad\ 9 \\
\indent{\textit{3.2. Gravitational instability}} \dotfill\quad\ 11 \\
\indent{\textit{3.2. Core-accretion versus gravitational instability}} \dotfill\quad\ 12 \\
{\bf 4. Interior structure properties }\dotfill\quad\ 13 \\
\indent{\textit{4.1. Thermodynamic properties}} \dotfill\quad\ 13 \\
\indent{\textit{4.2. Internal structure and composition}} \dotfill\quad\ 18 \\
\indent{\textit{4.3. Energy transport properties}} \dotfill\quad\ 19 \\
{\bf 5.  Atmospheric properties}\dotfill\quad\ 21 \\
\indent{\textit{5.1. Chemistry, Clouds, and Opacities}} \dotfill\quad\ 21 \\
\indent{\textit{5.2. Irradiation Effects}} \dotfill\quad\ 24 \\
\indent{\textit{5.3. Non-equilibrium Chemistry}} \dotfill\quad\ 27 \\
\indent{\textit{5.4. Biosignatures}} \dotfill\quad\ 28 \\
{\bf 6.  Evolutionary properties}\dotfill\quad\ 30 \\
\indent{\textit{6.1. Basics of evolutionary models}} \dotfill\quad\ 30 \\
\indent{\textit{6.2. Initial conditions}} \dotfill\quad\ 30 \\
\indent{\textit{6.3. Cooling and contraction history}} \dotfill\quad\ 31 \\
\indent{\textit{6.4. Mass-radius relationship}} \dotfill\quad\  33 \\
{\bf 7.  Star-planet interaction. Tidal effects}\dotfill\quad\ 34  \\
\indent{\textit{7.1. Star-planet orbital parameters}} \dotfill\quad\ 34 \\
\indent{\textit{7.2. Orbital evolution}} \dotfill\quad\  36 \\
\indent{\textit{7.3. Tidal energy dissipation}} \dotfill\quad\  41 \\
{\bf 8.  Observational constraints}\dotfill\quad\  42 \\
\indent{\textit{8.1. The radius anomaly of transiting exoplanets}} \dotfill\quad\  42 \\
\indent{\textit{8.2. Brown dwarfs versus massive giant planets }} \dotfill\quad\ 47  \\
\indent{\textit{8.3. Hot Neptunes and evaporation process}} \dotfill\quad\  49 \\
\indent{\textit{8.4. Light of extra-solar planets}} \dotfill\quad\ 50  \\
{\bf 9.  The future}\dotfill\quad\ 53  \\
{\bf Acknowledgments}\dotfill\quad\ 56  \\
{\bf References}\dotfill\quad\ 56 \\
\newpage

\section{\label{intro}Introduction}

The end of the twentieth century saw a revolution in our knowledge of planetary systems. The discovery of the first extrasolar planet in 1995 marked the beginning of a modern era and a change of our perception of planets. The discoveries continue apace and reveal an extraordinary diversity of planetary systems and exoplanet
physical properties, raising new questions in the field of planetary Science.
More than 400 exoplanets have now been unveiled by radial velocity measurements,
microlensing experiments and photometric transit observations.  They span a wide range of masses from a few Earth masses to a few tens of Jupiter masses (\cite{udry07}). The realm of terrestrial exoplanets starts to open its doors with the lightest known exoplanet, GJ 581e,
detected by radial velocity and having a mass $M. \sin i = 1.9$ Earth masses ($\mearth$) (\cite{mayor09}).
At the opposite end of the mass range,
an ambiguity appears on the nature of newly discovered objects.  
It now becomes clear that giant planets and brown dwarfs  share  a common mass range, likely between  a few Jupiter masses and several tens of Jupiter masses. 

Coincidently, the first brown dwarf was discovered at nearly the same time as the first exoplanet, but with different technologies and observational strategies. 
Identifying these two astrophysical body populations remains an open issue. Brown dwarfs are supposed to form like stars through the gravitational fragmentation of a molecular cloud
while planets are thought to form in a protoplanetary disk subsequently to star formation. A brown dwarf is an object unable to sustain stable hydrogen fusion
because of  the onset of electron degeneracy  in its central region. 
This definition provides a theoretical upper limit for their mass: objects below $\sim$ 0.07 solar masses ($\msol$) or 70 Jupiter masses ($\mjup$) belong to the brown dwarf realm (\cite{chabrier00}).
But brown dwarfs with increasingly small masses have now begun to be discovered, reaching masses characteristic of our solar system gaseous giants. In parallel, planet hunters have discovered  massive objects around  central stars with orbital properties 
characteristic of planetary systems.
They have been faced with an unprecedented difficulty to name their favourite object
and the community even called into question the definition of a planet (\cite{basri06}). This uncertainty about 
how to call these objects yields some sterile, semantic  debates which shed more confusion than light. 

On a theoretical point of view, the physical properties of giant planets and brown dwarfs
can be described within similar theoretical frameworks, the two families  of objects being closely related
in terms of atmospheric and interior properties. Because of different formation processes,
distinctions are expected concerning their composition and content of heavy elements and these differences must be taken into account in theoretical models. 
But clearly,
 the nascent theory for exoplanets has inherited from our knowledge 
of the Solar System planets, mostly developed during the past century, and from the recent 
progress performed in the modelling of brown dwarfs. Therefore, the description of the physical properties
of exoplanets is built  on a combination of our knowledge in planetary
and stellar science.

The aim of our review is to present the status of the modern theory describing 
atmospheric,
interior and evolutionary properties of exoplanets, as well as their formation mechanisms. 
Most exoplanets yet detected through
radial velocity and photometric transits are giant planets, characterised by the presence of
a gaseous envelope  mostly made of hydrogen and helium (\cite{udry07}). Crucial constraints on their
structure are revealed by photometric transit and Doppler follow-up techniques, which provide
a measure of their mass and radius.  Information on the mean density and bulk composition of several exoplanets is thus now available, drastically extending the knowledge
of planetary structures restricted till recently to our four giant planets. 
Atmospheric properties of exoplanets also start to be measured and first constraints on temperature
structure, composition and dynamics are now available. 
At dawn of their discovery, no observational constraints on Earth-like exoplanets are yet available.
Our review is thus essentially devoted to giant planets and we will only  briefly describe  the first theoretical efforts devoted to the description of terrestrial exoplanets.

The review is organised as follows. As introductory sections to the field, \S \ref{observations} and \S \ref{section_formation} provide brief overviews of solar and extra-solar system planets and of our current understanding of planet formation, respectively. Interior structure properties of terrestrial
to jovian planets are described in \S \ref{section_interior}, while \S \ref{section_atmosphere} is devoted to their atmospheric properties. Evolutionary properties, describing planet cooling and contraction history and the mass-radius relationship are presented in \S \ref{section_evolution}. Tidal effects and star-planet interactions are examined in \S \ref{section_tide}. Current observational constraints ({\it e.g} transiting radii, planetary light detection) are analysed in \S \ref{constraints}. Finally, some future advancements expected in the field are discussed in \S \ref{future}.

\section{\label{observations} A brief overview of observations}

In this section, we present observed properties of Solar  and extra-solar planets 
which are relevant for the understanding of exoplanet physical properties.
For more details, we invite the reader to refer to  the reviews 
by \cite{guillot05, guillot08}  on 
Solar System giant planets and by \cite{udry07} on statistical properties of exoplanets.

\subsection{\label{section_SS} Lessons from our Solar System}

The understanding of planetary structure starts with the extensive works 
conducted on our Solar System giant planets. Important constraints 
on  interior structures of our four giant planets are provided by measurement of their gravity field
through analysis of the trajectories of the space missions Voyager and Pioneer. Our giant planets
are fast rotators, with rotation periods of about 10 hours for Jupiter and Saturn, and about 17 hours
for Neptune and Uranus. Rotation modifies the internal structure of a fluid body and yields departure of 
the gravitational potential  from a spherically symmetric potential. Within the framework of a perturbation theory largely developed for rotating stars and planets, sometimes referred to as the theory of figures (\cite{zharkov78}), the gravitational
potential can be expressed in terms of even (for axysymetric bodies) Legendre Polynomials and gravitational
 moments $J_{\rm 2n}$. The latter are related to the inner density profile
 of the rotating object. Measurements of $J_2$, $J_4$ and $J_6$ for the four giant planets yield stringent
 constraints on their inner density profile. An abundant literature exists on the application of the theory of figures to our four giant planets and most models are based on the so-called three-layer model (\cite{stevenson82}). For Jupiter and Saturn, models  assume
  that the planetary interior consists of a central 
 rocky and/or icy core\footnote[1]{The term "rock" usually refers to silicates (Mg-, Si- and O-rich compounds)
 and the term "ice" involves a mixture of volatiles dominantly composed of water, with traces of
 methane and ammonia.}, an inner ionised helium and hydrogen envelope, and an outer neutral He and molecular H$_2$ envelope (see \cite{guillot04, guillot05} and references therein). Table 
\ref{table1} summarizes the results of a detailed analysis performed by \cite{saumon04}, taking into account uncertainties on the equation of state for hydrogen at high density. More recent models for Jupiter were derived by 
\cite{militzer08, nettelmann08} based on improved
equations of state for H and He derived from first-principles methods. They however reach contradictory
conclusions. While \cite{nettelmann08} essentially agree with the results of \cite{saumon04},
\cite{militzer08} find a larger core, of about 16 $\mearth$, and exclude a solution without a core. 
These two recent works illustrate the remaining uncertainties on planetary modelling and on equations of state of matter under 
conditions characteristic of giant planet interiors (see \S \ref{section_EOS}).  

\begin{table}
\caption{\label{table1}Interior properties of Jupiter and Saturn according to \cite{saumon04}.
$M_{\rm core}$ is the mass of the rocky/icy core; $M_{\rm Z}$ the mass
of heavy elements in the envelope; $M^{\rm tot}_{\rm Z}$ = $M_{\rm core}
+ M_{\rm Z}$ the total mass of heavy elements; $Z/Z_\odot$ is the ratio
of heavy elements in the planet to that in the Sun.}

\begin{tabular}{ccc}
\hline\noalign{\smallskip}
  & Jupiter & Saturn\\ 
  & (317.8 $\mearth$) & (95.1 $\mearth$)\\
\noalign{\smallskip}
\hline\noalign{\smallskip}
$M_{\rm core}$ & 0 - 11 $M_\oplus$ & 9 - 22 $M_\oplus$\\
$M_{\rm Z}$ & 1 - 39 $M_\oplus$ & 1 - 8$M_\oplus$ \\
$M^{\rm tot}_{\rm Z}$ & 8 - 39 $M_\oplus$ & 13 - 28 $M_\oplus$\\
$Z/Z_\odot$ & 1 - 6 & 6 - 14\\ 
\hline
\end{tabular}
\end{table}

The lighter giant planets, Uranus and Neptune, are more enriched in heavy elements
than their massive companions.
A wide variety of models can match the mass/radius of these planets.
Three-layer models with a central rocky core, an ice layer and an outer H/He envelope 
suggest
an overall composition of 25\% by mass of rocks, 60-70\% of ices and 5-15\% of gaseous H/He
(\cite{podolak91}). 
Other solutions exist, as suggested by \cite{podolak95},
assuming, instead of a pure ice second layer, a mixture
of ice, rock and gas (see Table \ref{table2}). According to recent
estimates, an upper limit for the H/He mass fraction is about 5 $\mearth$
for Uranus and 4.7 $\mearth$ for Neptune if only
rock and H/He are present (\cite{podolak00}).

\begin{table}[h!]
\caption{\label{table2} Examples of interior composition for 
Uranus and Neptune according to \cite{podolak95}. The models
assume (i) one rock layer, (ii) one layer composed of a mixture of rock,
ice and H/He and (iii) a third layer of H/He.
$M_{\rm rock}$, $M_{\rm ice}$ and  $M_{\rm H/He}$
are the total masses of rock, ice and H/He respectively.}
\begin{tabular}{ccc}
\hline\noalign{\smallskip}
  & Uranus & Neptune \\ 
  & (14.5 $\mearth$) & (17.1 $\mearth$)\\
\noalign{\smallskip}
\hline\noalign{\smallskip}
$M_{\rm rock}$ &  3.7 $M_\oplus$ & 4.2 $M_\oplus$\\
$M_{\rm ice}$ & 9.3 $M_\oplus$ & 10.7 $M_\oplus$ \\
$M_{\rm H/He}$ & 1.5 $M_\oplus$ & 2.2 $M_\oplus$ \\
\hline
\end{tabular}
\end{table}

Interestingly enough, models for Uranus assuming that each  layer is homogeneous
with an adiabatic temperature profile fail to reproduce the gravitational moments. An interesting solution
for this problem, suggested by \cite{podolak95}, is to assume that some parts of the layers
are not homogeneously mixed. In the presence of a persistent molecular weight gradient,
convection can be confined in numerous homogeneous layers separated by thin diffusive 
interfaces. This process, known as layered or double diffusive convection, provides a solution
to reproduce the observed gravitational moments in Uranus (\cite{podolak95}). We will see in 
section \ref{section_transit}  that this process may also be relevant for extra-solar transiting planets.

The analysis of the atmospheric composition of our giant planets also shows
a significant enrichment in heavy elements. Abundances of several elements (C, N, S, Ar, Kr, Xe)
have been measured in situ
by the Galileo probe for Jupiter and they show a global enrichment compared to solar values
of about a factor 3 (\cite{encrenaz05, guillot08}). For Saturn, spectroscopic detections of methane and ammonia
suggest significant enrichment of C (about a factor 6 \cite{encrenaz05}) and N (about a factor 2 \cite{guillot08}), although with large uncertainties for the latter element. For Uranus and Neptune, 
carbon is significantly enriched, with large  factors $>$ 20 (\cite{encrenaz05}) while the abundance of N/H is comparable to that of Jupiter and Saturn 
(\cite{guillot08}).

The interior and atmospheric properties of our giant planets bear
important consequences on our general perception of planetary  structure.
The observational evidences that our giant planets are enriched in heavy material support our general understanding of planet formation (see \S \ref{section_formation}) and guide the development of a general theory for exoplanets. 
 
 \subsection{\label{section_observation}Observed properties of exoplanets}
 
\begin{figure}
\begin{center}
\includegraphics[width=11cm]{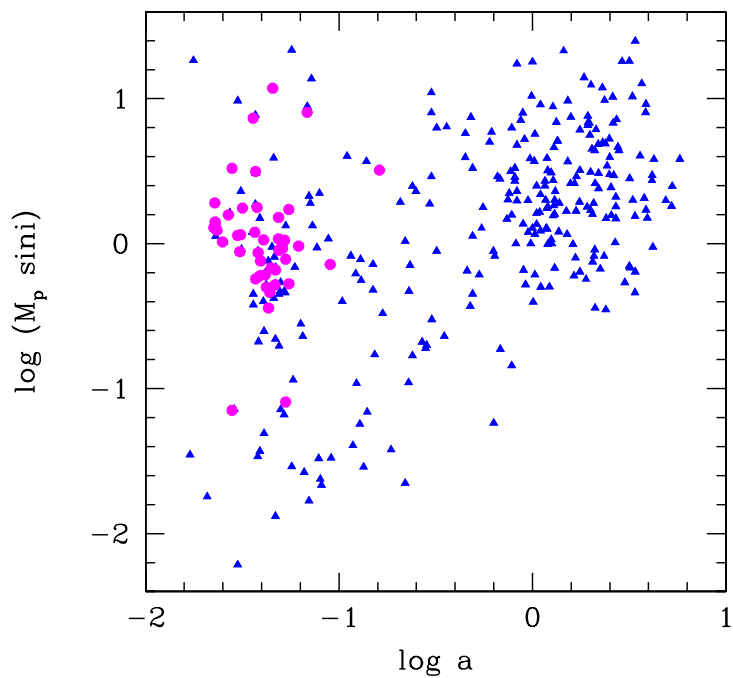}
\end{center}
\caption{Mass of known extra-solar planets (in $\mearth$) as a function of orbital distance (in AU). 
Triangles are planets detected by radial velocity surveys (from the web site of J. Schneider: http://exoplanet.eu/catalog-RV.php)
and solid circles are transiting planets (from the web site of F. Pont: http://www.inscience.ch/transits/).}
\label{mass_exoplanet}
\end{figure}

 The description of exoplanet physical properties must  encompass the wide
 variety of planetary masses and orbital separations yet discovered, as illustrated in Fig. \ref{mass_exoplanet}.  About 30\% of exoplanets have
 an orbital separation $a$  less than 0.1 AU.
 Irradiation effects from their parent star must thus be accounted for for a correct description
 of their structural and evolutionary properties. 
 Another compelling property of exo-planetary systems is the correlation between planet-host star metallicity\footnote{The metallicity is defined as the mass fraction of all chemical elements heavier than helium.} and frequency of planets. The probability of finding giant planets is  a strong function of the parent star metallicity, indicating that an environment enriched in heavy material favours planet formation  (\cite{udry07}).  This correlation, however, is not observed  for light Neptune-mass planets
  (\cite{udry07}), although the statistics is still poor. About twenty of these
 light planets have yet been discovered (\cite{udry07}). These trends provide clues about
 their formation process, as discussed in \S \ref{section_formation}. 

\begin{figure}
\begin{center}
\includegraphics[width=10cm]{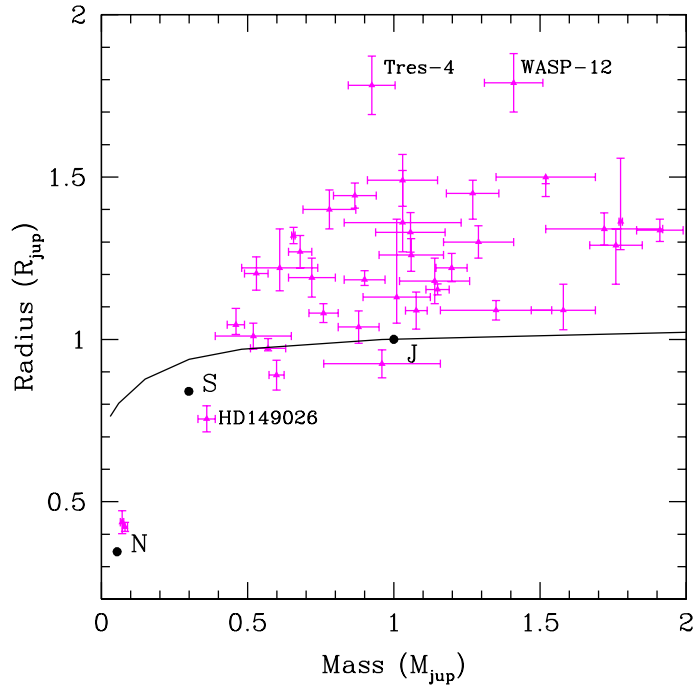}
\end{center}
\caption{Mass - radius diagram for transiting planets.  
The positions of Jupiter, Saturn and Neptune are indicated by full circles. The curve 
shows theoretical predictions for solar metallicity objects $Z=2$\% (models from \cite{baraffe08}). 
Note that the weak enrichment in heavy material of Jupiter ($Z \sim$ 10\%, see Table \ref{table1}) is counterbalanced by the effect of rotation (see \S \ref{section_SS}), and the radius of Jupiter is coincidentally reproduced by solar composition models. 
Data taken from the web site of F. Pont: http://www.inscience.ch/transits/.}
\label{transit_mr}
\end{figure}

  Crucial information on  interior structure and bulk composition
 of exoplanets are unveiled by  objects  transiting in front of their parent star. 
 About sixty of these planets
 have yet  been detected, revealing an extraordinary variety in mean planetary densities
 and composition. As illustrated in Fig. \ref{transit_mr},
 some exoplanets are significantly denser, thus more enriched in heavy material than our own giant planets. One of the most remarkable discovery is the Saturn mass planet HD 149026b with such a small radius that more than 70 $\mearth$ of heavy material is required to explain it (\cite{satos05}). This discovery raises the question of the
  formation process of planets with such a large amount of heavy material. More importantly,
 it tells us that exoplanets, like our Solar System planets, may contain large amounts of heavy material, supporting our current understanding of planetary formation (see \S \ref{section_formation}).
  
Another puzzling property revealed by Fig. \ref{transit_mr} is the abnormally large radius
of a significant fraction of transiting planets. Oddly enough, the very first transiting planet
 ever discovered, HD209458b (\cite{charbonneau00}), was found with a large radius which still challenges our understanding of planetary structure. At the time of this writing, the two most inflated exoplanets known are TRES-4b,
 with a radius $R_{\rm p} = 1.78 \rjup$ and a mass $M_{\rm p}$= 0.9 $\mjup$ (\cite{sozetti08})
 and WASP-12b, with $R_{\rm} = 1.79 \rjup$ and $M_{\rm p}$= 1.4 $\mjup$ (\cite{hebb08}). These observations indicate that new mechanisms are required to inflate close-in planets. Whether these mechanisms are peculiar, operating only under very specific conditions and
 planetary system configurations, or whether basic physics is missing in the modelling
 of close-in giant planets are still open questions. Several mechanisms have been proposed since the discovery of HD209458, but no firm answer has been obtained for now. In this review, we will present the main proposed mechanisms and discuss their status (\S \ref{section_transit}).

\section{Planet formation}
\label{section_formation}

\subsection{Core-accretion model}
\label{core-accretion}

The presently most widely accepted scenario for giant planet formation is the so-called core-accretion model \cite{Mizuno80, Pollack96}. In this model, solid cores grow oligarchically in the surrounding nebula by accreting small planetesimals\footnote{defined as solid objects present in protoplanetary disks.},  located within the protoplanet's zone of gravitational influence, called the feeding zone, which extends over a few Hills radii ($R_H=a(M_p/3M_\star)^{1/3}$, where $M_p$ and $M_\star$ denote the protoplanet and star mass, respectively  and $a$ the planet's orbital radius). The planetesimal accretion rate $dm_{s}/dt\propto \Sigma_s\approx 10^{-5}\mearth$/yr, where
$\Sigma_s$ denotes the local surface density of solids at the orbital radius ($\Sigma_s\simeq 3\,(a/5\,{\rm AU})^{-3/2}\,\gc2$
for the minimum-mass solar nebula (MMSN\footnote{The minimum-mass solar nebula is defined as  the protoplanetary disk of solar composition that contains the 
 minimum amount of solids necessary to build the planets of  our Solar System (\cite{weidenschilling77, hayashi81}).})). 
 
 Once the solid core has grown of the order of a Mars mass ($\sim 0.1\mearth)$,
it starts capturing an envelope of nebular gas and the protoplanet's growth is governed by
a quasistatic balance between radiative loss and accretion energy due dominantly to planetesimals (planetesimals are supposed to sink to
the planet's central regions whereas the accreted gas remains near the surface), with a negligible
contribution from the $PdV$ contraction work.
Both solid and gas accretion rates are relatively constant during this  phase, with gas accretion exceeding the planetesimal one. Above a critical mass $M_{crit}$, a
static envelope can no longer be supported. Gravitational contraction is now necessary to compensate the radiative
loss, which increases in turn gas accretion, leading to a runaway process, and the core accretes a massive gas envelope, becoming a newborn giant planet. For an envelope dust-dominated opacity ($\kappa \sim 0.1$-1 cm$^2$/g), $M_{crit}\sim 10\,\mearth$, although with large possible variations due to the sensitive
dependence of $M_{crit}$ upon the envelope composition (mean molecular weight, opacity), the planetesimal accretion rate and the distance of the planet to the star \cite{stevenson82, Rafikov06}. At this stage, the mass of accreted gas is comparable to
the mass of accreted planetesimals and the atmosphere has grown massive enough that its radiative loss can no longer be balanced by planetesimal accretion.
The massive (dominantly H/He) envelope can no longer maintain quasi-static equilibrium and it falls in free fall onto the central core. The nascent planet's radius is essentially fixed by this accretion shock condition. This process, and thus the expected radius and luminosity of young planets, however, remains ill determined, given the lack of a proper treatment of the radiative shock (see discussion in \S \ref{initial}, \cite{chabrier07, marley07}). 

The main problem faced by the conventional core-accretion model is that core growth takes longer than typical protoplanetary disk lifetimes, $\lesssim$ a few Myr \cite{Haisch01}. Giant planet cores can be obtained within the appropriate timescale either by increasing severely the disk density compared to the MMSN or by reducing drastically the accreting envelope opacity,
allowing rapid core contraction. A reduced opacity implies dust grains significantly larger than in the ISM, and thus efficient settling in the warmer parts of the disk, where they are destroyed \cite{Hubicky05, Podolak03}. 
Spiral density waves generated in the gas by the core, however, cause this latter to
migrate inward, the so-called type-I migration, with a characteristic timescale $t_I=a/|{\dot a}|\sim
\Omega^{-1}({M_\star \over M_p})({M_\star\over \Sigma_{gas} a^2})({H\over r})^2\approx ({M_\star \over M_p})({r\over 1\,{\rm AU}})$ yr, for values appropriate to the MMSN, where $\Omega$ denotes the protoplanet's Keplerian angular velocity and $H$ the disk scale height \cite{Ward97, Ward97b, Tanaka02}. This timescale
is much shorter than the time required to build up a 10 $\mearth$ core in standard core accretion models. These calculations, however, assume that the disk is laminar (linear planet-disk perturbations). Simulations including the effect
of (MHD) turbulence in the disk show that the mean torque does not converge towards the value obtained in a laminar disk (at least for the duration of the simulations) and that the usual type I migration is disrupted by turbulence \cite{NelsonPapaloizou04, Nelson05}. Analysis of these
stochastic torques suggests that a planet can overcome type-I migration for
several orbital periods, $P=2\pi/\Omega$. Above a critical mass, of the order of 30 $\mearth$ at 5 AU (although with large
uncertainties) for typical protoplanetary disk conditions, the planet's gravitational tidal torque exceeds the viscous torque, eventually stopping the motion of the protoplanet. Deposition of angular momentum in
the planet's vicinity, due to shock and viscous dissipation, pushes material away from the planet,
clearing an annular gap in the disk \cite{LinPapaloizou86,Ward97}. The planet is now locked in the disk and undergoes type-II migration with a timescale given by the
disk viscous timescale $t_{II}\simeq 5\times 10^6\,\Omega^{-1}({10^{-4}\over \alpha})({10^{-1}\over (H/r)})^2 ({r\over 10\,{\rm AU}})^{3/2}$ yr, assuming a Shakura-Sunyaev type viscosity law, $\nu=\alpha c_s H$, where $c_s$ is the isothermal sound speed averaged over the vertical structure \cite{LinPapaloizou86,Ward97}. Note that this timescale is now independent of the protoplanet's mass. Gap clearing and disk dissipation will limit the accretion onto the central core once the nebula starts falling down around it - although further accretion is possible if the protoplanet has an eccentric orbit - and the planet's final mass is
set up by these limits. 

Taking into account
migration processes in the core-accretion model has been found to speed up core growth by increasing the supply of planetesimals, avoiding the depletion of the feeding zone obtained in the in-situ formation models and solving the timescale problem of the standard
core-accretion scenario \cite{Alibert05, IdaLin08}. Planets, including our own solar system giants, now form on a timescale consistent with disk lifetimes, with the appropriate observational signatures  
\cite{Alibert05}. These models, however, have to reduce the conventional type-I migration rate by a factor $\ge 10$. It is currently not clear whether turbulent-induced stochastic migration can yield such a decrease over
significant ($\gtrsim 10^5$ yr) timescales and the real impact of stochastic migration remains an open issue. Interestingly, recent core-accretion simulations including the concurrent growth and migration of {\it multiple} embryos find a global {\it negative} impact on planet formation: while migration
indeed extends the domain of accretion for an embryo, this latter must also compete with other earlier generation embryos which
have depleted the inner regions of solid material, reducing the final number of giant-planet cores 
\cite{Chambers08}. 

In summary, including migration in the conventional core-accretion model succeeds in forming giant planets within appropriate timescales down to the inner edge of the disk.
These models, however, use disk surface densities or dust-to-gas ratios
about 2-3 times larger than the MMSN (suggesting that giant planet cores are unlikely to form in a MMSN) and require adequate planetesimal sizes and/or
viscosity parameters $\alpha$ in order for giant planet cores to grow rapidly before type-I or type-II migration moves them into the star. Unfortunately, these parameters, which involve complex processes such as
grain growth/fragmentation or turbulent viscosity are very uncertain and can vary over orders of magnitude.

\subsection{Gravitational instability}\label{GI}

The alternative theory for giant planet formation is direct gravitational fragmentation and collapse of a protoplanetary disc, the so-called gravitational instability (GI) scenario \cite{Cameron78,Boss97},
originally suggested to circumvent the timescale problem
of the original core-accretion scenario. Instability to axisymmetric (ring-like) perturbations in a disk occurs when the Toomre's stability criterion is violated, i.e. $Q=c_s\kappa_e / \pi G \Sigma\sim T^{1/2}\Omega / \Sigma \sim {M_\star \over M_d}{H\over r}<1$ \cite{Toomre64}, where $\kappa_e$ is the epicyclic frequency at some point in the disk
(for Keplerian orbits, the orbital and epicyclic frequencies are nearly the same whenever the disk mass is small compared to the stellar mass),
$\Sigma \approx M_d/r^2$ is the surface density, $H\sim c_s/\Omega$ is the disk vertical scale height and
$M_d$ is the disc mass contained within radius $r$. According to the Toomre criterion,
the disk becomes unstable to its own gravity whenever the stabilizing influence of differential rotation or pressure  is insufficient.
Note that this criterion is governed by the local density and is thus a {\it local criterion} for fragmentation. In the nonlinear regime, {\it global} spiral 
waves develop for values of $Q\lesssim 1.3$-1.7 with a growth time of a few orbital periods. This solution involves spiral modes that either saturate at low-amplitude via mode coupling, leading to rapid non-local angular momentum redistribution restoring gravitational stability with no disk fragmentation \cite{Laughlinetal97}, or fragment the disk. Estimates of disk surface densities and the fact that
typically $H/r\sim 0.1$ for protostellar discs
indicate that disks with $M_d\lesssim 0.1\,M_\star$ 
are usually stable to GI. A disk, however, might become
unstable during its evolution, for instance during its formation, if mass builds up faster than it is accreted by the star or, at later times, if the outer 
part of the disk, where stellar radiation
is negligible, becomes sufficiently cool.
For fragmentation to occur, however,  the disk must cool quickly enough to avoid entering a self-regulated phase, i.e. the disk cooling time $t_{cool}$ must
satisfy the condition, often called Gammie criterion, for very-thin disks $t_{cool}<\xi \,\kappa_e^{-1}$, with $\xi\sim 3$  \cite{Gammie01}.
Indeed, the energy loss rate determines the effect of the instability : isothermal disks, in which energy is easily lost (and gained, in gas expansion) remain unstable and evolve violently whereas adiabatic disks tend to heat up and become more stable. 

Energy transport and dissipation processes are thus key issues to determine whether or not planets can form from gravitational instabilities in a disk.  A detailed analysis of these conditions \cite{Rafikov05} shows that planet formation by GIs can occur only
 in very massive disks, $M_d\gtrsim 0.1\,\msol$, at the very upper end of the observed distribution, and at large orbital distances, $a\gtrsim 100$ AU. Even when vigorous convection occurs, it does not lower $t_{cool}$ enough to lead to fragmentation \cite{Rafikov07, Boley07}. Note also that stellar irradiation tends to hamper fragmentation. 
 
 An other key issue for
 planet formation by GI is the fate of the fragments. Even if the disk cools fast enough for fragmentation to ocur, the fragments have to last long enough to contract into planet embryos before being disrupted by tidal stresses, collisions or shocks. Moreover,
the typical mass scale associated with the fastest-growing density perturbations in a disc undergoing GI,  for
$H/r\sim 0.1$ and $M_\star=1\,\msol$, is of the order of a few Jupiter masses \cite{Rafikov05}. An important question is whether such a fragment can form a core since at such relatively high initial central temperatures, less than 1\%
of the gas can condense into grains and one has to invoke efficient sedimentation of silicate and iron grains to the center during the early contraction phase to be consistent with the enhanced heavy element abundances relative to solar values inferred for our giant planets (see \S \ref{section_SS} and Tables \ref{table1}-\ref{table2}). Therefore, it is far from clear that the peculiar composition and structure of our jovian planets can be explained within the GI model. 
  Fragmentation by GI thus remains controversial, with markedly different results from various groups, and requires that
the disk detailed thermal energetics are properly taken into account (see e.g. 
\cite{Durisen07, Dullemond08} for
recent reviews). 

\subsection{Core-accretion versus gravitational instability}

It is interesting to point out that disk instability predicts that even very young
($\lesssim 1$ Myr) stars should harbor gas giant planets, whereas the formation of such planets with
 the core accretion scenario requires a few Myr. Observational searches for the presence of genuine giant planets around $\sim$1 Myr-old stars will thus provide crucial tests for the two formation scenarios\footnote{For this test to be meaningful, it is crucial to identify the stellar companion as a genuine metal-enriched {\it planet}, not as a brown dwarf of similar mass. Indeed, GI-induced fragmentation can occur during the early (dynamical) stages of protostellar collapse and rapid disk accretion, but objects formed by
fragmentation in these cases are companion stars or brown dwarfs, not planets.}. 
As mentioned above, determination of the heavy element abundances of extrasolar planets
also provides a definitive test: compositional similarity of planets and their parent star would strongly favour gravitational collapse whereas significant heavy element enrichment of the planet with respect to the parent star composition would rule it out. The large heavy element enrichment inferred for
several transiting planets (see \S \ref{section_observation})
clearly supports the core accretion model. Last but not least, the efficiency of planet formation by GI should not depend on the disk metallicity, since gravitational collapse from the protoplanetary disk is a compositionally indiscriminating process, contrarily to the core-accretion scenario. The observed clear dependence of planet frequency with the host star metallicity (see \S \ref{section_observation})
and the suggested trend that metal-depleted stars seem to harbor lower mass planets ($M\,\sin(i) \lesssim 1\mjup$), i.e the lack of massive planets around metal-poor stars and the fact that stars hosting Neptune-mass planets
seem to have a flat metallicity distribution \cite{udry07} clearly suggest that metallicity plays a crucial role in planet formation, in agreement with the core-accretion model. 
Furthermore, statistical analysis of {\it null detections} in direct imaging surveys place 
constraints on the occurence of
giant extrasolar planets around FGKM stars. Calculations based on probability distributions
derived from observed mass and semi-major axis distributions of extrasolar planets
\cite{Cumming08}, show that 4 $\mjup$ and larger planets are found around less than 20$\%$ of stars 
beyond about 60 to 180 AUs, depending on the planet theoretical models, 
{\it i.e} beyond the equivalent of
extrasolar Kuiper belts, at 95$\%$ confidence level \cite{Lafreniere07, NielsenClose09}. 
Even though these calculations must be taken with caution (they do not include the few detected large orbit planets) and need to be confirmed by
more detailed statistical analysis, they suggest that
extrasolar giant planets at large separations ($\simgr 60$ AU) are rare. This observational constraint,
combined with the statistical conclusion that the less massive the star the 
lower the likelihood
to host a giant planet ($>$ 0.8 Mjup) \cite{Johnson07} and with the theoretical constraints
mentioned in \S \ref{GI}, strongly weakens the GI model and reinforces 
the core accretion one.
This latter thus appears
 most likely as the dominant scenario for planet formation. 
 
 The recently detected
3-$\mjup$ object Fomalhaut-b at projected separations $>100$ AU from the central star \cite{Kalas2008}, and other similar discoveries, challenge planet formation theories. Although local formation by GI is not excluded \cite{Boley09}, formation by core accretion at shorter orbital distances remains a possibility, the planets having migrated outwards under the action of either planet-planet scattering \cite{ScharfMenou09,Veras_etal09} or resonant interactions with an other planet in a common gap \cite{Crida_etal09}. GI-induced planet
formation, however, remains a plausible explanation in some peculiar situations, like for instance
for the planets recently discovered orbiting a
double SdB-M dwarf system \cite{lee08}. In that case, the circumbinary disk is likely to be massive enough
to become unstable.


\section{Interior structure properties}
\label{section_interior}

The impact of planetary internal compositions on their radius goes back to the pioneering works of Zapolsky \& Salpeter \cite{ZS69}, who considered various zero-temperature single element compositions, and to Stevenson \cite{stevenson82} for H/He, ice and rock planet compositions. Modern calculations, although basically similar to these studies, present improvements upon these works as they include modern equation of state (EOS) calculations, notably for H/He, and accurate atmospheric boundary conditions, taking into account irradiation from the parent star when necessary (see \S \ref{section_cooling}). 

\subsection{Thermodynamic properties }
\label{section_EOS} 

The mechanical structure and internal heat profile of a planet are entirely determined by the EOS of its chemical constituents. Indeed, due to efficient convective transport, the internal temperature profile is quasi-adiabatic.
In the pressure-temperature ($P$-$T$) domain
characteristic of planet interiors, elements 
go from a molecular or atomic state in the low-density outermost regions to an ionized, metallic one in the dense inner parts, covering the regime of {\it pressure}-dissociation and ionization. Interactions between molecules, atoms, ions and electrons are
dominant and degeneracy effects for the electrons play a crucial role, making the derivation of
an accurate EOS a challenging task. We examine below our present knowledge in this domain.

\subsubsection{Equation of state for H/He.}

The most widely used EOS to describe the thermodynamics properties of the gaseous H/He envelope of giant planets is the Saumon-Chabrier-VanHorn EOS (SCVH) \cite{SCVH95}. This semi-analytical EOS recovers numerical simulations and experimental results in the high-density and low-density regimes,
respectively, while, in its simplest form, interpolating over the pressure ionization regime. Thanks to the growth of computational performances, the properties of
dense hydrogen can now be calculated from first-principle or nearly first-principle quantum mechanical calculations. The last generation of these ab-initio calculations seem to converge towards an
agreement and to predict a substantially less compressible EOS for hydrogen, i.e. predict a lower density for a given pressure, than SCVH in the pressure-ionization domain, $P\sim 0.5$-4 Mbar
(see \cite{CSP06} for a recent review of these EOS models). This bears significant impact on the
internal structure of giant planets, in particular the size of the central core. A less compressible planet will tolerate
less heavy material for a given radius or, conversely, will have a larger radius for the same internal composition \cite{saumon04}. 

\begin{figure}
\begin{center}
\includegraphics[width=12cm]{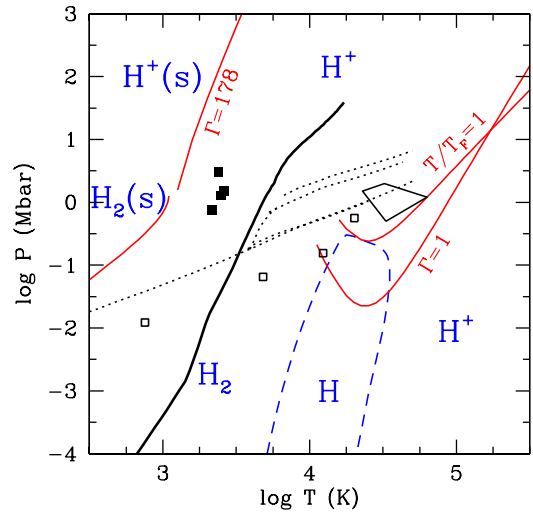}
\end{center}
\caption{Phase diagram of hydrogen. Solid lines indicate dimensionless physical parameters ($\Gamma=(Ze)^2/kTa_i$, where $a_i$ is the mean interionic distance, is the so-called plasma parameter,
with $\Gamma=1$ delineating weakly correlated from strongly correlated plasma regions and $\Gamma=178$ indicating the solid-liquid melting line for a bcc-lattice, and $\theta=kT/kT_F$, where $kT_F$ is the electron fluid Fermi energy, is the
so-called degeneracy parameter). The dashed curves indicate the 50\% dissociation and ionization boundaries. The dotted lines indicate regions probed by single, double and triple shock experiments on {\it deuterium} while filled squares show recent near entropic compression data \cite{Fortov}. The open squares
indicate single and double shock data for {\it helium} \cite{Nellis} while the small polygone indicates the locus of recent shock wave experiments from pre-compressed initial states \cite{Eggert08}.
The heavy solid line illustrates the isentrope of Jupiter. {\it Figure kindly provided by D. Saumon}}
\label{Hdiagram}
\end{figure}

Interestingly enough,
high-pressure experiments on fluid deuterium\footnote{Because of the higher density of deuterium compared with hydrogen, higher
shock pressures can be achieved experimentally for a given impactor speed.} or helium are now able to reach pressures and temperatures typical of the giant planet interiors ($P\gtrsim  1$ Mbar, $T\gtrsim  10^4$ K), probing the EOS in its most uncertain pressure-range. These experiments are dynamical, i.e. based on shock-driven compression. For a given initial state of the sample, mass, energy and momentum conservation across the shock, as given by the Rankine-Hugoniot relations, force the family of shocked states, which correspond to different shock velocities, to follow a Hugoniot curve in the $(P,\rho,T)$ phase diagram.
Dissociation or ionization processes absorb the corresponding amounts of energy and thus yield high degrees of compression with a modest temperature increase, whereas in the absence of such processes the energy of the shock is expended mostly in the kinetic degrees of freedom with a corresponding increase of temperature, following a different Hugoniot for the same initial state. Figure \ref{Hdiagram} portrays the phase diagram of hydrogen, with the pressure range presently probed by
high-pressure dynamical experiments on H and He, together with Jupiter's internal adiabat.
Various experimental techniques, however, give different results, with a $\sim 30$-$50\%$ difference in $P(\rho)$ in the maximum compression region for deuterium, $\sim 0.5$-1.5 Mbar, although 
the most
recent experiments seem to converge towards the "stiff" (least compressible) Hugoniot result \cite{Knudson03,Boriskov05}, in agreement with the aforementioned quantum mechanical simulations (see \cite{CSP06} for various theory-experiment comparisons). 
This issue must be settled before hydrogen pressure ionization can be considered as correctly understood.  

For the less explored case of liquid helium,
recent high-pressure experiments have been achieved up to 2 Mbar for various Hugoniot initial conditions, allowing to test the EOS over a relatively broad range. These experiments show a larger compressibility than for hydrogen, due to electronic excitations \cite{Eggert08} and are
in good agreement with the SCVH EOS while ab-initio calculations \cite{Militzer06} underestimate the compressibility. Clearly more of these experiments, exploring the so-called "warm dense matter" domain, are needed to fully assess the validity of the various EOS models, with a
crucial impact on our knowledge of the structure of Jovian planets \cite{saumon04,nettelmann08,militzer08} (see \S \ref{section_SS}).

The combined interactions of H and He in the {\it mixture} increase drastically the degree of complexity in
the characterization of the plasma. Not only the interactions between the two fluids will affect the
regime of pressure ionization compared with the pure components, but partial immiscibility between the two species has been suggested
to explain Saturn's excess luminosity for the age of the Solar System \cite{stevenson77,SS77,FortneyHubbard04},
and may occur inside some exoplanets. A phase separation in the interior of a
planet strongly affects its cooling history. Indeed, if H/He phase separation (or immiscibility of other species with H)
occurs in a planet's interior, He-rich dropplets will form in a surrounding H-rich fluid and will sink to the planet's center under the action of the gravity field. This extra
source of gravitational energy converted into heat slows down the planet's cooling, or alternatively yields a larger luminosity for a given age. The strongest suggestion for a phase separation in Jupiter or Saturn interiors
comes from the observed depletion of helium in their atmosphere, with a mass fraction $Y=0.238\pm0.05$ measured by the Galileo probe for Jupiter, and a more uncertain determination,  $Y=0.18$-0.25 inferred from Voyager observations for Saturn, to be compared with the protosolar nebula value, $Y=0.275$ \cite{guillot08}. 
Unfortunately, given the aforementioned difficulty in modelling the
properties of H or He alone, and the necessity to simulate a large enough number of particles for
the minor species (He in the present case) to obtain
statistically converged results, no reliable calculation of the H/He phase diagram can be claimed so far.
As mentioned above,
pressure-ionization of pure hydrogen and helium must first be fully mastered before the reliability
of the calculations exploring the
behaviour of the mixture can be unambiguously assessed\footnote{Note that the recent claim \cite{StixrudeJeanloz} that H/He phase separation can not take place in Jupiter's interior, because metallisation of He should occur at lower pressure than previously expected, is not correct. Although such a facilitated metallization (by itself a model-dependent result)
could exclude the suggested H$^+$-He immiscibility \cite{Stevenson79}, it does not preclude at all the H$^+$-He$^{++}$ one.}.

Resolving these important issues concerning the H and He EOS must await (i) unambiguous experimental confirmation of  the H and He EOS at high pressure, (ii) unambiguous confirmation
of the reliability of the theoretical calculations, in particular in the pressure ionization regime, (iii) guidance from experiments to predict the behaviour of the H/He mixture under planetary interior conditions. Progress
both on the experimental and theoretical side will hopefully enable us to fullfill these criteria within the
coming years (see section \ref{future}).

\subsubsection{Equation of state for heavy elements.}
\label{eosz}

According to the composition of the protosolar nebula, the next most abundant constituents after hydrogen and helium in gaseous giant planets, but the most
abundant ones in ice giants and Earth-like planets, consist of C, N and O, often refered to as "ices" under their molecule-bearing volatile forms (H$_2$O, the most abundant of these elements for
solar C/O and N/O ratios, CH$_4$, NH$_3$, CO, N$_2$ and possibly CO$_2$). The remaining constituents consist of silicates (Mg, Si and O-rich material) and iron (as
mixtures of more refractory elements under the form of metal, oxyde, sulfide or substituting for Mg in
the silicates). The behaviour of these different elements as a function of pressure, under the conditions typical of giant planet interiors
is not or poorly known. At
very high pressure, the categorizations of gas, ice and rock become meaningless and these elements
should become a mixture of closed-shell ions. The most
widely used EOS models for such elements are ANEOS \cite{ThomsonLauson72} and SESAME \cite{LyonJohnson92}, which describe the thermodynamic properties of water, "rocks" (olivine (fosterite Mg$_2$SiO$_4$) or dunite in ANEOS, a mixture of silicates and other heavy elements called "drysand" in SESAME) and iron. These EOS consist of
interpolations between existing Hugoniot data at low to moderately high
($\lesssim 0.5$ Mbar) pressure and Thomas-Fermi or more sophisticated first-principle calculations at very
high density ($P\gtrsim 100$ Mbar), where ionized species dominate. Interpolation, however, provides no insight about the correct structural and electronic properties of
the element as a function of pressure, and thus no information about its compressibility, ionization stage (thus conductibility), or even its
phase change, solid or liquid. All these properties can have a large impact on
the internal structure and the evolution of the planets. A detailed comparison between these EOS, and the impact of the uncertainties on the radius determination for Neptune-like and Jupiter-like planets has been conducted in \cite{baraffe08}. The largest difference between the various EOS models, reaching up to $\sim 40$-60$\%$ in $P(\rho)$ and $\sim 10$-15$\%$ on the entropy $S(P,T)$,
occurs in the $T\sim10^3$-$10^4$ K, $P\sim10^{-2}$-$1$ Mbar interpolated region, the typical
domain of Neptune-like planets \cite{baraffe08}. For these objects, such an uncertainty on the heavy element EOS translates into a $\sim 10\%$ uncertainty in the radius after 1 Gyr,
and to larger uncertainties at earlier ages (Fig. 3 of \cite{baraffe08}).

For solids, the lattice structure energy dominates
the thermal vibration (phonons) contribution and, once the planet becomes cool enough for the core to
become solid, the thermal contribution of this latter to the cooling of the planet can be neglected. This is
not true at higher temperature, i.e. during the earlier stages of the planet's evolution, when the
heavy elements are predicted to be in a liquid state, according to their EOS. In that case, thermal effects can substantially modify the zero-temperature structure contribution and assuming a zero-T or low-T core, i.e. neglecting the core release
of thermal energy $(dU/dt)_{\rm core}$ and gravitational energy ($P {dV \over dt})_{\rm core}$ can drastically underestimate its true gravothermal contribution, affecting the general cooling of the planet. The thermal contribution of the heavy elements to the cooling of the planet must then properly be taken into account for a correct description of the planet's evolution \cite{baraffe08}. 

\subsection{Internal structure and composition}

There is compelling evidence from our own solar system planets that they are substantially enriched in
heavy elements (C,N,O), with a $\sim 3$ to $\sim 6$ times solar value for Jupiter and Saturn, respectively, and even larger contrasts for Uranus and Neptune (see \S \ref{section_SS}), as expected from a formation by core-accretion. This should apply as well to the
discovered exoplanets. Indeed, the small radii of some of the observed transiting planets can only be reproduced if these objects
are substantially enriched in heavy material, with a total of several tens to hundreds of Earth masses.
For all these objects, including our own Jovian planets, however, uncertainties remain on (i) the total amount of heavy material, (ii) the
respective fractions of "rocks" (silicates) and "ices", and (iii) the distribution of this heavy material in the
planet's interior. As mentioned above, the first uncertainty (i) stems primarily from the present uncertainties on the various EOSs and should decrease with further experimental and theoretical progress.
Addressing the two other issues necessitate to differentiate gaseous from solid/liquid planets.

\subsubsection{Earth-like to super-Earth planets ($\lesssim 10\mearth$)}
\label{eos_earth}

As discussed in \S \ref{section_formation}, planets below about 10 $\mearth$, usually denominated as super-Earth down to Earth-like planets, are not massive enough to enter the unstable, run-away regime leading
to rapid accretion of a large gaseous envelope onto the central core. Post-formation degassing or oxydization processes can only produce a teneous gaseous atmosphere, with no significant consequences for the planet's contraction. Therefore, these objects consist essentially of solids or liquids rather than gases, making their structure determination from mass-radius observations less uncertain than for more massive
planets (see below). The mass-radius relationship for these low-mass planets has been parametrized as $R=R_{ref}(M/\mearth)^\beta$, with $R_{ref}=(1+0.56\,\alpha)R_\oplus$ and
$\beta=0.262(1-0.138\,\alpha$), for the rocky or ocean super-Earth planets in the mass range 1-10 $\mearth$ (\cite{valencia07}), where $\alpha$ denotes the water mass fraction, and $\beta=0.3$ for planets between 10$^{-2}$ to 1 $\mearth$,
with a weak dependence upon the iron to silicate ratio Fe/Si \cite{sotin07}. Note that incompressible (constant density)  material corresponds to $\beta=1/3$. These parametrizations appear to be rather robust, despite the uncertainties in the EOS and in the iron/silicate fraction
\cite{fortney07,seager07,sotin07}. This provides a sound diagnostic for transiting Earth-like planet detections and the possible identification of the so-called "ocean planets", planets composed dominantly of water \cite{Kuchner03,leger04}, as opposed to the terrestrial (Fe-rich) planets. As mentioned in
\ref{eosz}, current uncertainties in
the high-pressure behaviour of silicates, ices and iron alloys prevent more precise information such
as the detailed internal composition or the size and the nature, solid or liquid, of the central core.
Exploration of the iron phase diagram and melting curve, in particular, a subject of prime importance for the characterization of
the Earth inner core, has become even more interesting since the discovery of exoplanets of a few Earth masses (see \S \ref{intro}) and the expected wealth of transiting Earth-size planets with the Kepler mission (\cite{borucki03}). 
While dynamic (Hugoniot) experiments produce too large temperatures at the pressure of interest for the Earth,
they might become relevant for the super-Earth objects. 
This again points the need for high-pressure experiments for planetary materials in the appropriate
$\gtrsim 10^3$ K, $>$ Mbar regime. 

\subsubsection{Neptune-like to super-Jupiter planets ($\gtrsim 10\mearth$)}
\label{eos_jupiter}

For planets with a $\gtrsim 10\%$ by mass gaseous (H/He) envelope, this latter
essentially governs the gravothermal evolution of the planet. For instance, a 10$\mearth$ planet retaining a modest 10\% H/He envelope is 50\% larger
than its pure icy counterpart \cite{Adams08,baraffe08}. Under such conditions, a variety of internal compositions, either water-rich or iron-rich can produce the same mass-radius signature \cite{Adams08,baraffe08}
and detailed information about the planet's internal structure, other than inferring its bulk properties,
becomes elusive.
One of the main uncertainties about the internal structure is that we do not know whether these heavy elements are predominantly concentrated into a central core or
are distributed more or less homogeneously throughout the gaseous H/He envelope. 

A summary of the main consequences of the uncertainties (i) in the EOS,
(ii) in the chemical composition and (iii) in the
distribution of the heavy elements for planets with masses $\gtrsim 10\mearth$ can be found in
\cite{baraffe08}. In particular, \cite{baraffe08} show that for a global metal enrichment $Z\gtrsim 15\%$, all heavy material being either gathered in a core or
 distributed homogeneously throughout the envelope yields a $\gtrsim 10\%$ difference in radius after 1 Gyr for Neptune-mass planets and  a $\gtrsim 4\%$ difference for Jovian-mass planets.

\subsection{\label{section_transport} Energy transport properties}

\subsubsection{Giant planets}

The large radiative opacity of planetary material yields completely  inefficient heat transport by photons, except in the most outer layers close to the planet photosphere (see \S \ref{section_atmosphere}). Transport by conduction, resulting from collisions 
during random motion of particles, may in some cases be relevant. In  the central 
part of H/He dominated planets, thermal conductivity is dominated by 
electronic transport with conductivity $\kappa_{\rm T} \sim 10^{-1}$ cm$^2$ s$^{-1}$ (\cite{stevenson77}). If no electrons are available, as in the outer envelope, 
conductive transport is dominated by the less efficient molecular motions with
thermal conductivity 
$\kappa_{\rm T} \sim 10^{-2}$ cm$^2$ s$^{-1}$ (\cite{stevenson77}). Conduction by electrons (or eventually phonons) may also dominate in central cores composed of heavy material (see \cite{baraffe08}).  But the dominant energy transport mechanism in giant planet interiors is convection.
Large scale convection is extremely efficient in transporting
heat (\cite{hubbard84, stevenson85}) and the temperature profile is close to adiabatic.
Superadiabaticity is extremely small, with $\nabla_{\rm T} - \nabla_{\rm ad}  \simle 10^{-8} $ in most of the  interior\footnote{$\nabla_{\rm T}=d\, lnT/d \, ln P$ is the local temperature gradient and $\nabla_{\rm ad}=(dlnT/dlnP)_S$ the adiabatic gradient.}. For H/He dominated planets, convective velocities derived from 
 the mixing length formalism vary between $\sim$ 10 cm s$^{-1}$, in the central regions,
  and a few 100 cm s$^{-1}$ in the outer layers. 
  
As initially stressed by \cite{stevenson85}, it is a conventional assumption in giant planet modelling to postulate that their interiors are fully convective and thus homogeneously mixed. This assumption has never been proven to be valid and has been questioned in the case of our own giant planets by \cite{stevenson85}. As mentioned in \S \ref{section_SS}, more recent
 models for Uranus have also suggested the possible existence of non-homogeneous
 regions where convection is not efficient. If compositional gradients exist, convection can break into convective layers separated by thin diffusive layers, the so-called double-diffusive or layered convection, or, if layers do not persist, overstable modes of convection can lead to the growth of small-scale fluid oscillations, the so-called oscillatory convection, becoming more alike an enhanced diffusion process than a large scale convective process \cite{stevenson79, SS77}. 
 Layered convection is known to occur in laboratory experiments and in some regions of the Earth's oceans or great lakes. In these conditions, the heat flux transport  is significantly reduced because of the presence of multiple diffusive layers. If this process is present under conditions characteristic of giant planet interiors,
 heat transport in the diffusive layers is due to conduction with 
  above-mentioned thermal conductivities  (\cite{cb07}). Because diffusion limits heat transport,
 the internal heat flow of the planet is significantly reduced compared with 
 that of a fully convective object. This may bear important consequences on the planet
 structure and evolution, as discussed in \S \ref{section_transit},
 slowing down its global cooling and contraction (\cite{stevenson85, cb07}).

\subsubsection{\label{section_earth} Terrestrial exoplanets}

Many efforts are devoted to the modelling of massive terrestrial planets,
or super-Earths, 
essentially composed of heavy material, with a negligible amount of gas.
Simplified models have been developed in order to investigate a wide range of compositions and masses (\cite{seager07, fortney07}). These models involve simple internal structures and compositions
(see \S \ref{eos_earth}).
 They use zero- or uniform, low temperature equations of state for the heavy material and thus do not consider the planet's thermal structure. Such simplified models are advocated on the basis that, so far, only exoplanet bulk compositions can be inferred from mass and radius measurements, so that
 detailed planet interior models are not necessary \cite{seager07}. In the future, however, internal structure models reaching a degree of sophistication comparable to our own Earth's description might become required for terrestrial exoplanets. In this perspective, detailed models of
massive Earth or super-Earth analogs have been developed (\cite{sotin07,valencia06,grasset09}). They rely on our knowledge of the Earth's internal structure, appropriately
rescaled to super-Earth masses. They involve layered internal structures with different compositions and phase changes,
including an iron-rich core, lower and upper mantles composed of silicates
and an outer water layer (icy and/or liquid), as well as the related temperature profiles. The effect of large amounts of water, characterising the so-called "ocean-planets",  on the mass-radius relationships has also been investigated by \cite{leger04}. 
The thermal profile is constructed according to 
the Earth thermal structure, which is determined by convective transport in each layer
and is characterised by important variations at the interfaces. According to \cite{valencia06}, the thermal structure has little effect on the radius of an Earth-like planet but imposes conditions on the compositional phases, particularly for water
at the surface. This is of prime importance for the characterisation of ocean planets, in order to know
 whether or not they can have a liquid water surface \cite{leger04,valencia07}. 
Typical terrestrial geological processes such as plate tectonics are now introduced
in some exo-Earth models, with the claim that this mode of convective transport mechanism
might be important for massive terrestrial planets (\cite{valencia07b}). 

 
\section{\label{section_atmosphere} Atmospheric properties}

The wide variety of planets both in and out of our solar system provide
excellent laboratories for understanding how atmospheres evolve with time and
react to their environments.   Atmospheres by themselves hold a nearly endless
supply of complex and interesting physical problems, but are also the primary
link between observations and theory.  The majority of observational techniques
for studying planets involve capturing photons that have emerged from the
atmosphere, placing extreme importance on our ability to successfully model
atmospheric behavior.   The grouping and location of substellar mass objects
(brown dwarfs and giant planets) on color-color and color-magnitude diagrams is
largely due to the sculpting of the emergent spectrum by atmospheric opacity
sources.   Also, as already mentioned above, the atmosphere regulates the
release of energy from the interior and establishes the upper boundary
condition for interior models.  In the following sections, we
discuss several of the important aspects of atmosphere modelling.

\subsection{Chemistry, Clouds, and Opacities \label{clouds}}

Chemistry is at the heart of most atmospheric phenomena and establishes the
distribution of the elements among various compounds.  Among these compounds
are the primary opacity sources which in turn play key roles in the radiative
transfer.   Planetary atmospheres are commonly assumed to be ideal gasses in
chemical equilibrium.  These assumptions greatly simplify the determination of
mole-fractions (or partial pressures) of most compounds that, under these
assumptions, depend only on temperature and pressure.  
 
In principle, chemical equilibrium models are extremely simple and require the
solution of a set of coupled nonlinear equations (for mass and charge
conservation and including either equilibrium constants or chemical potentials).
There are, however, practical considerations such as obtaining the necessary
thermochemical data (specific heat, entropy, enthalpy, etc.) of enough
compounds to ultimately end up with a realistic picture of the ensemble of
chemical species.  One must also correctly handle phase-transitions 
that produce liquid and solid species, which may or may not remain aloft in the
atmosphere.  Some of the first chemical equilibrium models applied to gas giant
atmospheres were carried out for Jupiter and Saturn \cite{Lewis1969,
FegleyLodders1994}. Later, with the discovery of brown dwarfs, a variety of
chemical equilibrium models were explored \cite{Tsuji1996, BurrowsSharp1999}.

\begin{figure}
\begin{center}
\includegraphics[width=10cm]{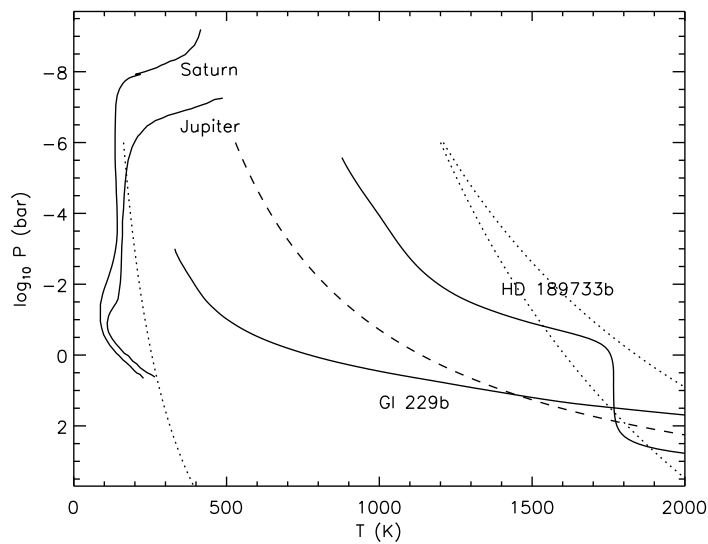}
\end{center}
\caption{Temperature versus pressure for the atmospheres of Saturn, Jupiter,
the brown dwarf Gl 229B, and the hot-Jupiter HD189733b.  The dotted lines are
the condensation curves of H$_2$O(ice), MgSiO$_3$ and Fe (from left to right).
Equal mole fractions of CO and CH$_4$ follow the dashed line, with CO favored
at higher temperatures.}
\label{struc}
\end{figure}

\begin{figure}
\begin{center}
\includegraphics[width=11cm]{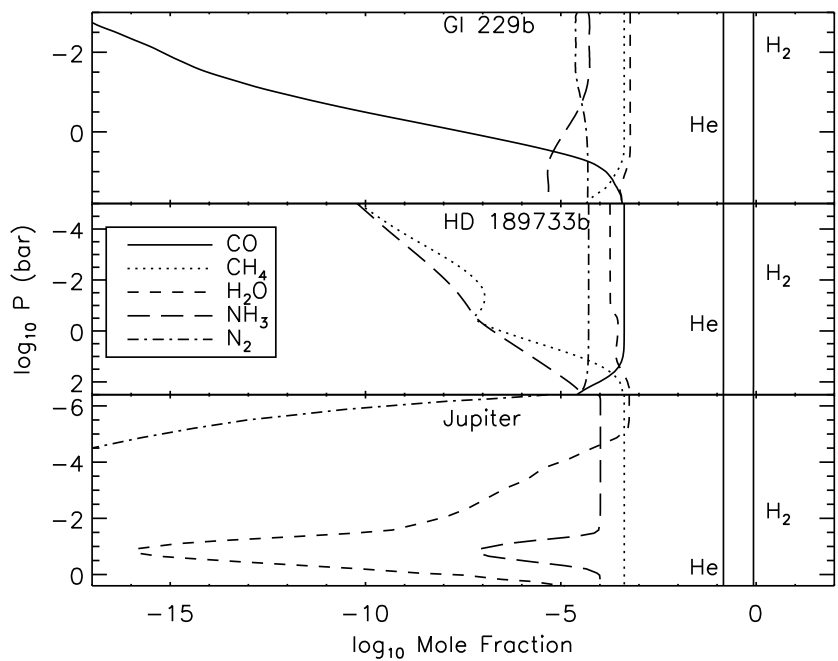}
\end{center}
\caption{Equilibrium mole-fractions with pressure (a proxy for height) for the three temperature-pressure profiles shown in Fig. \ref{struc} and for the most abundant gas phase species.  Drops in H$_2$O and NH$_3$ gas-phase mole-fractions are produced by the formation of water and ammonia ices. Departures
from these equilibrium values have been observed in Gl 229B and Jupiter, see discussion in text.}
\label{chem}
\end{figure}

The temperature structures for Jupiter, Saturn, a hot-Jupiter exoplanet (HD
189733) and a cool brown dwarf (of T spectral type) are shown in Fig. \ref{struc}.
In this figure pressure is used as a proxy for height, with pressure increasing
toward the center of the object.   For the two solar system planets the
temperature structures have been carefully measured, while the brown dwarf and
hot-Jupiter structures are theoretical predictions (while accommodating most available
observational constraints).  Figure \ref{chem} compares the mole fractions (or
equivalently the partial pressure divided by the total pressure) for several of
the most important molecules in a gas having the same element abundances as the
Sun.   Molecular hydrogen and He are by far the dominant constituents of the
gas since H and He are orders of magnitude more abundant in the Universe than
any other element. The  next three most abundant elements (C, N, and O) make up
the other dominant gas-phase molecules, H$_2$O, CO, CH$_4$, N$_2$, and NH$_3$.    From
Jupiter to a hot-Jupiter, the atmospheric temperature range is in the thousands
of Kelvin spanning a regime where water ice can form (in the case of Jupiter)
and where all the water is in the gas phase.  This broad temperature range also
leads to a role reversal for CO and CH$_4$ as the dominant carbon-bearing
molecule.   In Jupiter, equilibrium chemistry predicts essentially no CO and
nearly all of the carbon bound in CH$_4$, while the reverse is true for the
hot-Jupiter.  The brown dwarf has intermediate temperatures but is in the
region where water and methane are substantially more abundant than CO.
Ammonia begins to increase in concentration near the temperature found  in the
brown dwarf example, and becomes increasingly important in the very low
temperature atmospheres of Jupiter and the other solar system giant planets.  

At the low temperatures encountered in planetary atmospheres, phase transitions
can occur resulting in formation of liquid and solid ``condensates".   The
formation of condensates substantially alters the overall gas phase mole
fractions and, for example, explains the fairly dry conditions in Jupiter's
atmosphere as water is expected to condense into ice with water clouds deep in
the atmosphere ($\sim 5$ bar).  At higher temperatures other perhaps less familiar condensates
are expected to form involving silicate grains and many other minerals (Fe,
MgSiO$_3$, Mg$_2$SiO$_4$, Al$_2$O$_3$ and so on). In the presence of gravity
and vertical/horizontal mixing, these condensates can potentially form stable
clouds or potentially sink completely out of the atmosphere, taking with them
elements that can no longer participate in the chemistry.  Including clouds in
atmosphere models is a significant ongoing challenge and a variety of
approaches are being explored.  Extreme limiting assumptions include either no
clouds at all or clouds in pure chemical equilibrium, resulting in thick
clouds extending high up in the atmosphere.  More sophisticated models are used
to explore the intermediate cases, many of which are phenomenological in
design, with the expectation that the important cloud properties can be
described by a manageable set of parameters with values ultimately determined
empirically.  The cloud-model parameters most frequently used control cloud
thickness, particle size, and particle distribution.  Alternatively, some
models attempt to follow the micro-physics of nucleation and grain growth
coupled to mixing within the gas.  So far, no single model provides a
comprehensive physical description of clouds.  There is a long history of cloud
modelling for the giant planets in our Solar system \cite{Rossow1978} and many
of these have inspired cloud models used for exoplanet atmospheres
\cite{Ackerman2001}.  Through the study of brown dwarf atmospheres, a great
deal has been learned about cloud formation across a wide variety of
low-temperature environments that can be directly applied to exoplanet
atmospheres.  For a review of competing cloud models, mostly applied to brown
dwarf atmospheres, see \cite{Helling2008}.


The importance of the chemistry and clouds are translated to the rest of the
model atmosphere via the opacities, either directly or indirectly.  Opacity
sources are crudely divided up as either continuous (e.g., scattering,
bound-free, or free-free), or line opacities (atomic and molecular transitions)
and the relative importance of these is highly dependent on temperature and
pressure as well as wavelength.   The continuous opacity sources are generally
well characterized in atmosphere models; however, for many grains, optical
constants have either incomplete wavelength coverage or are not available at all.
Perhaps the most problematic source of opacity over the years has been the
molecular line opacity, where either the line data were simply not available or
the ability to include all of the necessary lines was computationally
prohibitive (requiring straight-means, K-coefficients, or the ``Just
Overlapping Line Approximation" (JOLA) to be used. For a description of these,
see \cite{Allard1997}).  Modern computers are now
quite capable of include millions of spectral lines, and model atmospheres can
now include all the line data available to produce very realistic spectra. Some
of the biggest improvements have been made in the completeness of line data for
water vapor (\cite{AMES, barber06}). Other molecules like CH$_4$ still lack important line data at high
temperatures (that are now very important given the temperature ranges
encountered in planetary atmosphere studies).

\subsection{\label{irradiation} Irradiation Effects}

Of the more than 400 planets known, a third of these orbit their host star
within 0.1 AU and, thus, receive a substantial amount of energy from the star.
This large number of short period exoplanets is attributed to observational
selection effects but, nonetheless, has greatly broadened our expectations for
giant planets and in particular the conditions present in their atmospheres.
In most of these cases, the amount of {\em extrinsic} energy received by the
planet from the star greatly exceeds the energy leftover from the planet's
formation that slowly leaks from the interior.  Consequently, unlike longer
period planets and isolated brown dwarfs, the dominant energy source is a
function of the orbital separation and the spectral type of the host star and
is less dependent on an exoplanet's mass and age (except for very young
planets).  

The incident flux can be defined by the following equation, where $R_{\star}$
and $T_{\star}$ are the host star radius and effective temperature, and $\alpha \in [1/4,1]$
is a scaling factor used to crudely account for day-to-night energy redistribution.

\begin{eqnarray}
F_{inc} = \alpha\left(\frac{R_{\star}}{r(t)}\right)^2 \sigma T_{\star}^4.  
\end{eqnarray}

\noindent Since exoplanets are found with a wide range of eccentricities, the
general (time-dependent) planet-star separation is given by,

\begin{eqnarray}
r(t) = \frac{a(1 - e^2)}{1 + e\cos\theta(t)},
\end{eqnarray}

\noindent where $a$ is the semi-major axis, $e$ is the eccentricity, and
$\theta(t)$ is the angle swept out by the planet during an orbit.

The atmospheric temperature of isolated stars (and of widely separated planets
and brown dwarfs) is often described by a single characteristic temperature,
the so called effective temperature, $T_{eff}$, defined such that $\sigma
T_{eff}^4$ equals the total energy per unit area radiated from the surface.  A
similar characteristic temperature is useful for describing the temperature of
irradiated planets, however, it is customarily called the equilibrium
temperature ($T_{eq}$) to avoid confusion with the internal ($T_{int}$)
effective temperature which is a measure of the energy contribution from the
interior.  Since energy is conserved, a planet in equilibrium must reradiate
the energy it receives from the star along with the energy escaping the
planet's interior.  Thus the equilibrium effective temperature is given by,

\begin{eqnarray}
\sigma T_{eq}^4 = \sigma T_{int}^4 + (1 - A)F_{inc},
\end{eqnarray}

\noindent where $A$ is the bond albedo.

As one would expect, the extrinsic flux heats the dayside of giant planet
atmospheres to temperatures well above those found in our Solar System.  The
predicted temperature structure of HD189733b (Fig. \ref{struc}) illustrates
this point.  The intrinsic flux of this planet is very close to that of
Jupiter, while the predicted temperatures across the dayside atmosphere are
entirely maintained by stellar heating.  The discovery of the first
hot-exoplanet, 51 Peg b, inspired many predictions for the properties of
irradiated planetary atmospheres \cite{Saumon1996, Guillot1996, Seager1998,
Goukenleuque2000, Sudarsky2000} with each of these early works employing
simplifying assumptions ranging from ad hoc temperature structures, artificial
upper/lower boundary conditions, and sparse chemistry and opacity descriptions.
Despite simplifying assumptions, these early models successfully predicted many
of the fundamental properties now revealed by observations and paved the way
for more detailed models.  More sophisticated 1-D models including state-of-the
art chemical and radiative transfer simulations have been explored for a broad
range of hot-exoplanet atmospheres, and have demonstrated the diversity and
complexity one should expect in hot-Jupiter atmospheres, both chemically and
spectroscopically \cite{barman01, Sudarsky2003, Barman2005, Seager2005,
Fortney2006b, Burrows2008}.


Early on, it was clear that traditional 1-D model atmospheres faced a geometric
problem concerning the natural division of a hot-exoplanet into day and night
sides.  Many of the short-period planets should be synchronously rotating (see \S \ref{section_tide}),
ensuring that the planet is irradiated constantly on the same hemisphere.  This
of course was not a new problem as stellar and planetary atmosphere modelers
have been dealing with this issue for decades earlier. In many ways, the
situation is far more similar to what is often encountered in irradiated
stellar atmospheres since the expected day-night temperature differences for
hot-exoplanets greatly exceeds what one finds in the Solar System (assuming
hydrostatic and radiative-convective equilibrium).  The frequently used
approach is to model the two sides (day and night) separately and scale the
incident flux by a geometric factor ($\alpha$ in the equations above) to
account for the redistribution (or lack of) of 
absorbed stellar flux across either the day side or the entire atmosphere.  

The value of this global redistribution factor has only recently been estimated
using observations in a few cases and more sophisticated 3-D dynamical
simulations are needed to predict wind speeds and horizontal heat transport.
Early 3-D global circulation models \cite{showman02} found that wind speeds
could exceed 1 km s$^{-1}$ and that day-night temperature differences of 500K
were possible at photospheric depths (i.e. the spectrum forming region).
These authors also stressed that dynamics has consequences for the chemistry,
potentially leading to observable departures from equilibrium (see below).
Different approaches have been used to model the circulation patterns in
hot-exoplanets, including 2-D \cite{burkert05,Langton2007,Cho2008} and 3-D
\cite{Cooper2005, Cooper2006, Dobbs-Dixon2008, Showman2008} models, most of
these motivated by varying successes within the Solar system planets.  To
accommodate the computational expense of following multi-D global circulations,
these models make various levels of sacrifices when it comes to the radiative
transfer in the atmosphere models.  There still remains a substantial
disconnect between the 1-D static atmosphere models with sophisticated
radiative transfer and the multi-D global hydrodynamical models; however,
efforts are underway to bring the two together \cite{Showman2009}.  While these
circulations models differ in the details (some have large vortices like the
spots on Jupiter, while others have single dominant westward jets), most agree
that strongly irradiated planets will develop a small number of broad
flows/jets \cite{Menou03}.  This differs from the Solar System planets which have numerous
narrow jets.  See \cite{showman07} for a recent detailed review of exoplanet
circulation models.

\subsubsection{Albedo}

Observational upper limits on the reflected starlight of hot-Jupiters yield
a rather small value for the geometric albedo. None of the attempts to detect the reflected light
from the ground during secondary eclipses has been successful so far. Space-based photometric observations of HD209458b
at visible wavelengths are the most constraining upper limit so far, and give a value less than 0.17 at 99.5\% confidence level \cite{Rowe_etal08},
significantly smaller than Jupiter's value of 0.5. These observations suggest that most of the
incident starlight of a hot-Jupiter, at least for HD209458b-like conditions, is absorbed, ruling out
the presence of many reflective clouds in this type of object. Further observations on a large sample of transiting
planets are necessary in order to obtain more stringent information about the albedos and the
atmospheric properties of these objects.

\subsection{Non-equilibrium Chemistry}

Departures from equilibrium chemistry can occur for a variety of reasons, such
as non-local phenomena (external radiation) and time-dependent mixing (vertical
and/or horizontal).  For example it is believed that enhanced tropospheric CO
in Jupiter's atmosphere is due to rapid vertical mixing and very slow chemical
reaction timescales to convert CO back to CH$_4$.  Similar mixing-induced 
equilibrium departures are thought to occur in the atmosphere of brown dwarfs
(e.g. Gl 229B), impacting both the CO/CH$_4$ ratios and the N$_2$/NH$_3$ ratios. 

The (net) chemical reactions responsible for converting CO to CH$_4$ and N$_2$
to NH$_3$ are,
\begin{eqnarray}
CO + 3H_2 \stackrel{slow}{\longrightarrow} CH_4 +H_2O,
\end{eqnarray}
and
\begin{eqnarray}
N_2 + 3H_2 \stackrel{slow}{\longrightarrow} 2NH_3.
\end{eqnarray}
\noindent The individual times ($\tau_{chem}$) for the two reactions
above are very long, from left to right, at low temperatures and pressures
(often greater than 10$^6$ years) compared to vertical upwelling, which
operates on year timescales or less.  In both planetary and brown dwarf
atmospheres the true mixing timescales ($\tau_{mix}$) in the radiative layers
are highly uncertain and often approximated as $\tau_{mix} \sim {H^2}/K_{zz}$,
where $H$ is the pressure scale height and $K_{zz}$ is the eddy diffusion
coefficient -- a free parameter ranging from 10$^2$ to 10$^5$ cm$^2$ s$^{-1}$ (see
\cite{Griffith1999} for more details).  When $\tau_{mix} >> \tau_{chem}$, the
chemistry is expected to be in equilibrium.  All other reaction pathways
involving important opacity sources occur very rapidly and, thus, are not
expected to be perturbed from equilibrium by mixing.  See \cite{Griffith1999,
Saumon2003, Saumon2006} for applications of this procedure to brown dwarf
atmospheres and \cite{Bezard2002} for CO in Jupiter's atmosphere. 

As already mentioned above, global circulations can lead to horizontal/vertical
mixing and cause departures from equilibrium chemistry.  In many of the
hot-Jupiters large mole fractions of CO can persist even in cool regions ($P
\simle 1$bar)  where chemical equilibrium predicts CH$_4$ as the dominant
carbon bearing molecule in a manner very similar to that described above for
brown dwarf atmospheres \cite{Cooper2006}.  Looking back at Fig. \ref{chem},
mixing is expected to elevate the CO mole fractions at low pressures ($P \simle
1$ bar) by many orders of magnitude above the chemical equilibrium curves
shown.  NH$_3$ could also be an order of magnitude below the equilibrium value
through much of the atmospheres shown.  Photochemistry driven by external
irradiation has been studied extensively in Solar System planets (see
\cite{Moses2000} for a review) but has only been studied in close-in giant exoplanets for a few specific cases \cite{Liang2004}. 
Also, a new field  develops with the study of photochemistry of Earth-like atmospheres,
one of the motivation being the analysis of "false positive" signals of life
due to photochemistry (\cite{selsis02,selsis08}, see below).

\subsection{Biosignatures}

\begin{figure}
\begin{center}
\includegraphics[width=10cm]{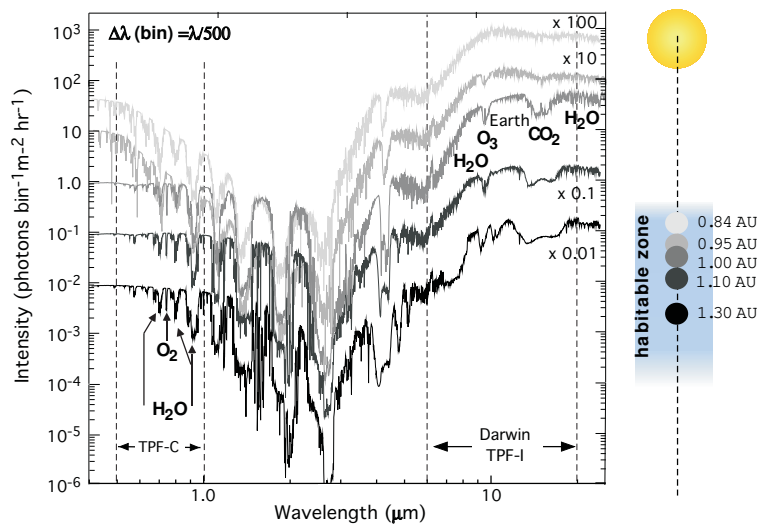}
\end{center}
\caption{Spectra of an Earth-like planet at different orbital distances from a Sun-like
star. The figure illustrates the evolution of the spectral signature of H$_2$O, O$_3$ and CO$_2$ as a function of
the planet's location in the habitable zone (\cite{selsis08}). The fluxes correspond to a system located
at 10 pc. Note that, for clarity, the intensity of each spectrum  at different orbital distances is multiplied by 
a factor given on the right hand side of the figure. {\it Figure kindly provided by F. Selsis}. }
\label{bio}
\end{figure}

\begin{figure}
\begin{center}
\includegraphics[width=10cm]{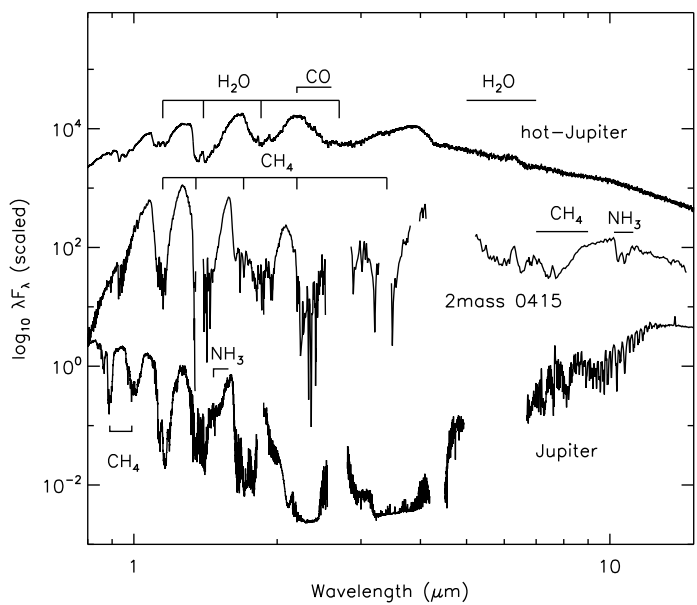}
\end{center}
\caption{Observed spectra for Jupiter (Rayner, Cushing \& Vacca (in prep.)
and \cite{Kunde2004}) and the brown dwarf 2M0415 \cite{Cushing2006} are compared
to a model spectrum of hot-Jupiter HD189733b \cite{Barman2008}. Figure adapted
from \cite{Cushing2006, Marley2008}.}
\label{specs}
\end{figure}

No Earth analogs have been yet detected, but their search is
one of the major goal of  ground- and space-based research programs of
the coming decades. Our current understanding of planet formation suggests
that terrestrial planet formation should be an efficient process. We thus expect
these planets to be common, as illustrated by our Solar System which has
three such planets (Earth, Mars and Venus). The search for signatures of life
on exo-Earths is one of the main motivations for these programs and certainly one of the most exciting scientific inquiry of the beginning of the century. A huge activity is now devoted to  astrobiology and we will only mention in this review the most basic biosignatures currently suggested.  Biogeochemical activity on a planet could manifest itself 
 through spectral features of the atmosphere. Current  search strategies, as derived
 for DARWIN/TPF (\cite{cockell08}), are thus 
 based on the spectroscopic detection of compounds that could not be present on a planet in the absence of life. The search for biomarkers 
is based on the assumption that extraterrestrial life shares fundamental characteristics with life on Earth. This later is based on carbon chemistry and requires liquid water as solvent. Other paths for life, based on a different chemistry, could perhaps exist but
the signatures of the resulting life-forms are so far unknown.  The need for liquid water leads to the concept of Habitable 
Zone defined as the region around a star where the surface temperature of
Earth-like planets allows the presence of liquid water (see Fig. \ref{bio}). This zone depends on the stellar luminosity and thus evolves in time with the star.  Its definition also depends on complex
processes on the planet, such as the concentration of greenhouse gases or
geological activity (see \cite{gaidos07, selsis08} and references therein). 
Any biomarker should  include the signature of H$_2$O, which is 
 a requisite for "earth-like" life.
The presence of H$_2$O, O$_2$
(or O$_3$) and CH$_4$, NH$_3$ or CO$_2$ would imply some biological activity,
so that these elements are considered as favourite biomakers (\cite{gaidos07, cockell08}). However, as pointed by \cite{selsis02}, the unique detection of one of these compounds may be ambiguous. Indeed, O$_2$, and hence O$_3$, can be produced by  photochemistry. The combined detection of O$_2$ with H$_2$O and CO$_2$, which are  important for habitability, would however provide a robust signature of biological photosynthesis (\cite{selsis02, gaidos07,cockell08}).
Similarly, the presence of CH$_4$ or NH$_3$ together with O$_2$ or O$_3$ would be  good biomarkers (see \cite{cockell08} and reference therein), as demonstrated by the observations of the Galileo probe, as it passed near the Earth and detected simultaneously O$_2$ and CH$_4$ (\cite{sagan93}). Note that methane and ammonia are not expected
to be abiotically produced on habitable, Earth-size planets, in contrast to a common
production in cold hydrogen-rich atmospheres of giant planets 
(see \S \ref{clouds} and Fig. \ref{specs}). The analysis of biological spectral signatures on an exoplanet is thus optimised if its physical properties, such as its mass and radius, can be also determined. 

\section{\label{section_evolution} Evolutionary properties }

\subsection{Basics of evolutionary models}

Planetary evolution is described  by the standard conservation equations, written
in a spherically symmetric configuration and  widely used
in stellar evolution calculations (\cite{guillot05}). The complete evolutionary problem requires the
coupling between  interior (see \S \ref{section_interior}) and atmospheric (see \S \ref{section_atmosphere}) structures.
The rate at which internal heat escapes depends on the surface properties,
and thus on the outer boundary conditions connecting interior and atmospheric structures. Modern models for planets incorporate
realistic atmospheric boundary conditions using frequency dependent atmosphere
codes as described in  \S \ref{section_atmosphere}.  This is required for a correct description of the thermal profile and effective temperature of objects having cold
molecular atmospheres (\cite{chabrier00, hubbard02}). The inner and outer temperature-profiles are connected at large optical depths, where either the atmosphere becomes fully convective or radiative transport only involves Rosseland mean opacity, which is a weighted mean over all frequencies of the inverse monochromatic opacity $\kappa_\nu$ using the temperature derivative of the Planck distribution as weighting function (see \cite{mihalas78}). 
 The connection
is  done either at fixed pressure, usually at a few bars (\cite{burrows03, guillot05, fortney03}) or at fixed optical depth, usually at $\tau_{\rm Rosseland}$ = 100 
(\cite{chabrier00}). The numerical 
radius, corresponding to the outer boundary condition, provides
to an excellent approximation the planet's photospheric radius, 
where the bulk of the flux escapes ($\tau_{\rm Rosseland} \sim 1$). This stems from the negligible atmospheric
extension between the photosphere and the 
depth characteristic of the outer boundary conditions.

\subsection{\label{initial} Initial conditions}

Planet evolution is characterised by the release of gravitational  and internal energy from an initial (unknown !) entropy state.
The usual procedure, similar to the brown dwarf
and stellar cases, assumes a high initial entropy state, i.e. a large initial radius and luminosity \cite{burrows97, hubbard02, baraffe03}. Since this implies a relatively small ($\lesssim Myr$) Kelvin-Helmholtz timescale, this yields a rapid early evolution so that the initial conditions are forgotten within a few Myr and do not influence the subsequent evolution \cite{baraffe02}. The assumption of a hot initial state for planets has recently been questioned by \cite{marley07}. Using initial conditions derived from the core-accretion model, they find lower initial
entropy states than aforementioned. These authors thus suggest that young giant planets should be fainter and smaller than predicted by standard evolutionary models, and thus fainter and smaller than young brown dwarfs at the same age.
Although based on the planet embryo's core-accretion history rather than on a totally arbitrary initial condition, this approach, however, still suffers from the lack of a proper treatment of the final accretion shock, which determines the nascent planet's initial energy content and radius (see \S \ref{core-accretion}). The results of  \cite{marley07} are a direct consequence of the assumption that the accreting gas loses most of its internal entropy through the shock. Such assumption is derived from the shock boundary conditions of \cite{stahler80}, which were initially developed for the study of accretion onto protostars. Multi-D radiative transfer and hydrodynamical simulations of the accretion process and the resulting shock could provide more rigorous post-shock initial conditions for  young planet models.  Although still very challenging, this level of complexity seems to be required in order to improve the field.
Therefore, the determination of the planet initial conditions remains so far an unsolved issue, as discussed in \cite{chabrier07,alibert09}. 

Such a high sensitivity of early planet evolution to initial conditions, as previously stressed for low mass stars and brown dwarfs \cite{baraffe02},
has major consequences on the identification of planetary mass objects 
in young clusters and on detection strategies of future projects such as SPHERE and
Gemini Planet Imager. 
The work of \cite{marley07} also raises the question whether the faintness of young planetary mass objects may
be used as a criterion to distinguish a brown dwarf from a planet. The answer is non trivial but one must keep in mind that accretion through a disk (circumstellar or circumplanetary) is a common process of both star and planet formation. Moreover,  the conclusion that young planets should be faint relies on a treatment of accretion which can also be applied
to the formation of protostars and proto-brown dwarfs (see \cite{alibert09} and \S \ref{section_formation}). Thus, for the same reasons, brown dwarfs may as well be fainter
than predicted by current evolutionary models.

\subsection{\label{section_cooling} Cooling and contraction history}

\begin{figure}
\begin{center}
\includegraphics[width=10cm]{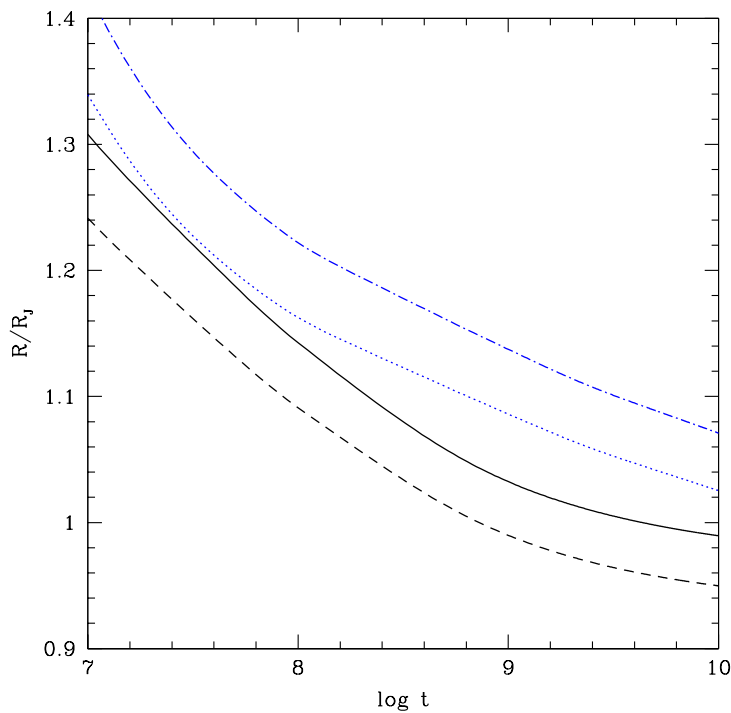}
\end{center}
\caption{Effect of irradiation and of heavy element enrichment on the 
radius $R$ as a function of time of a 1 $\mjup$ planet. Time is in year. Solid line: no core and no irradiation; dash-dotted line:
irradiation from a Sun at 0.045 AU, no core; dashed line: no irradiation and
a central core of water of 10\% of the planet's mass ($\mcore=31.8 \mearth$);
dotted line: irradiation and $\mcore=31.8 \mearth$.
(Models from \cite{baraffe08}). }
\label{tr_mj1}
\end{figure}

A planet contracts and cools down during its entire life on a characteristic thermal timescale $\tau_{\rm KH}  \sim {G M^2 \over RL}$. 
For a 1 $\mjup$ gaseous planet,
$\tau_{\rm KH} \sim 10^7$ yr at the beginning of its evolution, starting from
a hot initial state with luminosity $L > 10^{28}$ erg s$^{-1}$, and $\tau_{\rm KH} > 10^{10}$ yr after 1 Gyr, reaching luminosities $L \simle 10^{25}$  erg s$^{-1}$ (see \cite{burrows97, baraffe03}). As mentioned in \S \ref{section_observation}, a non-negligible
 fraction of exoplanets are at close distance from their parent star and their
 evolution is affected by irradiation effects. These effects are accounted for through
 the coupling of interior and {\it irradiated} atmosphere structures (see 
 \S \ref{section_atmosphere}). The heating of the outer layers by the incident stellar flux
yields an isothermal layer between the top of the convective zone
and the region where the stellar flux is absorbed. The top of the convective
zone is displaced toward larger depths, compared to the non-irradiated
case. The main effect is to reduce the heat 
loss from the planet's
interior, which can maintain higher entropy for longer time.
Consequently, the gravitational contraction of an irradiated planet
is slowed down compared to the non-irradiated case and the upshot
is a larger radius at a given age. This effect is illustrated in  Fig. \ref{tr_mj1}
on the evolution of a 1 $\mjup$ planet.

For Saturn and Jupiter mass planets with orbital distances  0.02-0.045 AU,
the typical effect of irradiation on the radius is about 10\%-20\% (\cite{baraffe03, burrows03, chabrier04, fortney07}). Since 
the total binding energy and the intrinsic luminosity of a planet decreases with its mass,  the lighter the planet, the smaller the ratio of its intrinsic flux to a given incident stellar flux. Consequently, the evolution and radius of a Neptune-like or smaller planet will be significantly more altered by irradiation effects, compared to the non-irradiated counterpart,
than for a more massive planet at a given orbital distance (see \S \ref{section_mr} and Fig. \ref{mr_theory}). 
Irradiation effects are thus quantitatively important and must be accounted for in the modern theory of exoplanet evolution. 

Furthermore, a consistent comparison between theoretical radius and observed
transit radius requires a subtlety due to the thickness of the planet atmosphere (\cite{baraffe03, burrows03, burrows07}). The measured radius
is a transit radius at a given wavelength, usually in the optical, 
which involves atmospheric layers above the photosphere.
Atmospheric extension due to  heating of the incident stellar flux
can be significant, yielding a measured radius larger than the 
theoretical or photospheric radius.
This effect can add a few \% (up to 10\%)
 to the measured radius (\cite{baraffe03, burrows03}). 


Inspection of transiting object mean densities suggests that some exoplanets, like the giant planets
of our Solar System, are enriched in heavy material. Models devoted to the analysis
of their properties must then account for such enrichment. Many efforts are now devoted
to the construction of models, covering a wide range of masses,  including
(i) different amounts of heavy elements
and (ii) different heavy material compositions (\cite{guillot06, burrows07, fortney07, seager07,
baraffe08}). Exploration of effect (ii)
is limited by available equations of state valid under conditions encountered
in planetary interiors (see \S \ref{eosz}). The most commonly considered materials are water, rock and iron.
For sake of simplicity, and given the currently large uncertainties on EOSs, on
the nature of the heavy elements and on their distribution inside the planet, current models
often assume that heavy materials are all contained in a central core. 
A detailed analysis of
the main uncertainties in current planetary models, due to EOS, composition and distribution of heavy material within the planet has been conducted in \cite{baraffe08} 
(see \S \ref {eos_jupiter}).
As shown in Fig. \ref{tr_mj1}, the radius
of a heavy material enriched planet  is smaller, at a given time, compared to the H/He gaseous counterpart.
Figure \ref{tr_mj1} also  illustrates the competitive effects of irradiation,
which increases $R$,
and of the presence of heavy material, which diminishes $R$.

\subsection{\label{section_mr} Mass-radius relationship}

\begin{figure}
\begin{center}
\includegraphics[width=10cm]{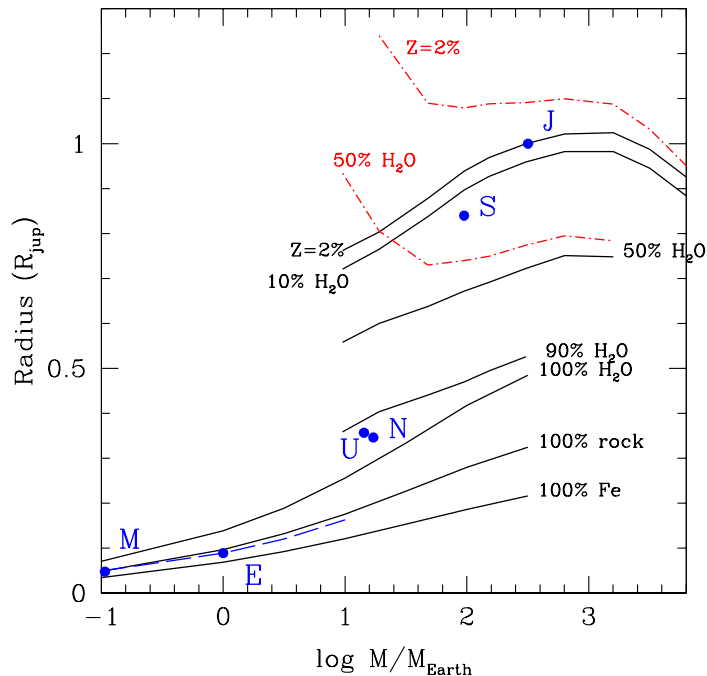}
\end{center}
\caption{Planetary radii at 4.5 Gyr as a function of mass, from 0.1 $\mearth$ to 20 $\mjup$. Models with solar metallicity ($Z=2$\%) and with different amounts of heavy material (water, rock or iron) are shown (Models from \cite{fortney07, baraffe08}). 
The "rock" composition here is olivine or dunite,  {\it i.e} Mg$_2$SiO$_4$. Solid lines
are for non-irradiated models. Dash-dotted curves correspond to
irradiated models at 0.045 AU from a Sun.  The long-dashed curve, between
0.1 $\mearth$ and 10 $\mearth$, is the mass-radius relationship for terrestrial planets, with detailed structures and compositions characteristic of the Earth, after \cite{valencia07}.
The positions of Mars, the Earth, Uranus, Neptune, Saturn and Jupiter are indicated by solid points. Note that pure heavy material planets (water, rock, iron) with masses $>$ 100
 $\mearth$ are unrealistic and are only shown for
illustrative purposes.}
\label{mr_theory}
\end{figure}

The mass-radius relationship of a planet at a given time of its evolution is entirely
determined by the thermodynamic properties of its internal constituents and its ability to transport and evacuate its internal entropy content. Figure \ref{mr_theory} portrays
the behavior of the mass-radius relationship of Earth-like to Jupiter-like planets.
The essential physics characteristic of the shape of the relation was described
in the pioneer work by \cite{ZS69}. Two competitive effects yield the well-known
flattening in the $M-R$ relationship around a few $\mjup$. In the low mass
regime, down to Earth masses, the dominant electrostatic contribution from the classical ions yields a relation $R \propto M^{+1/3}$, characteristic of incompressible
matter. As mass and density increase, the effects of partially
degenerate electrons start to dominate  over Coulomb effects, yielding a reversal
of the $M-R$ relation. Consequently, a maximum radius exists at a critical mass
which depends on the planet composition. \cite{ZS69} found a critical mass of 2.6 $\mjup$, corresponding to a maximum radius of $\sim$ 1 $\rjup$, for a gaseous H/He planet under the assumption of zero-temperature plasma. More recent calculations, based on improved EOS as described in \S \ref{section_EOS},
yield a critical mass $\sim 3 \mjup$. 

Figure \ref{mr_theory} also illustrates the increasing effect of
irradiation on the planetary radius as mass decreases for planets retaining a substantial atmosphere (see \S \ref{section_cooling}). The radius of
 a 20 $\mearth$ planet with 50\% H$_2$O is enhanced by $\sim$ 35\% if  irradiated by a Sun at 0.045 AU (\cite{baraffe08}). For terrestrial planets, the mass-radius relationship derived from detailed models assuming the same complex composition as for the Earth (\cite{valencia07}, see \S \ref{eos_earth}) are shown by the long-dashed line in Fig. \ref{mr_theory}. As discussed in \S \ref{section_earth}, these predictions are extremely close to the one derived from more simple models, assuming 100\% rock (\cite{fortney07}). 
 High accuracy (less than 5\%) on radius measurement, and to a lesser extent on mass measurement, is thus required to infer the detailed composition of a super-Earth exoplanet (\cite{sotin07, seager07,valencia07, grasset09}).  As shown by \cite{grasset09}, if the uncertainty on the radius of a super-Earth planet  is less than 3\%,
 the amount of water can be determined with an accuracy of less than 13\%.
 
 \section{Star-planet interaction. Tidal effects}
\label{section_tide}

 \subsection{Star-planet orbital parameters.}
\label{tide-int}

The geometry of an orbiting star-planet system of respective masses $ M_{\star}$ and $M_p$, illustrating the formation and dynamical evolution of the
system, is encapsulated in three main parameters: the semi-major axis $a$, associated with the mean orbital motion, $n=2\pi/P\simeq (GM_\star/a^3)^{1/2}$ where $P$ is the orbital period and $G$ the gravitational constant,
the eccentricity $e$ and the stellar obliquity $\epsilon$, defined as the angle between the angular momentum vectors of the planetary orbit and the stellar rotation axis. This latter quantity, which determines the spin-orbit angle, is an interesting
diagnostic for inferring the dominant interaction mechanisms at play in a protoplanetary disk: close spin-orbit alignment (as is the case for our solar system) is expected from quiescent tidal interactions or migration processes of a planet within the disk whereas planet scattering events are prone to misalignments of planetary orbit angular momentum and stellar spins \cite{Chatterjee08}. The inclination angle of the orbit relative to the sky plane is usually denoted $ i$ so that the line-of-sight projected value of a quantity retains a $sin \,i$ indetermination. A transit observation implies that $i \sim 90^0$. 
Although the stellar obliquity is generally not measurable, Doppler shift observations on the parent star throughout a transit offer the possibility to determine the angle between the sky projections of the two angular momentum vectors, i.e. the projected spin-orbit angle $\lambda$, which gives a lower limit of the true three-dimensional spin-orbit angle\footnote{Since the orbit of transiting planets is nearly edge-on, the angle $\lambda$ is related to the true stellar obliquity $\epsilon$ by $\cos \epsilon \simeq \cos \lambda \sin i_\star$, where $i_\star$ is the inclination of the stellar rotation
axis relative to the sky plane \cite{Winn_etal05}. Note that for transiting systems, the projected spin-orbit alignment angle can be zero
while the stellar obliquity is $90^0$ if the angle $i_\star$ between the stellar rotation
and the line-of-sight is also zero.}. The angle $\lambda$ is determined through the so-called Rossiter-McLaughlin (RM) spectroscopic effect \cite{Rossiter,McLaughlin}, originally applied to eclipsing binaries and recently extended to transiting planets \cite{Queloz00,Ohta05, fabrycky09} (see \cite{fabrycky09,Winn08} for recent reviews on RM observations).
To date, the RM effect has been measured for 12 transiting systems. For nine of these systems, the determinations are consistent with a small ($\lesssim 3^o$) or even zero-value of $\lambda$, although with significant error bars in some cases, indicative of well-aligned spin and orbit, similar to the solar system, even though one should keep in mind that these observations provide minimum (projected) values of the true stellar obliquity, as mentioned above. Alignment of the stellar spin and the planetary orbital axis is a strong confirmation of
planets forming in
a spinning protoplanetary disk surrounding the central protostar. 
Spin-orbit {\it misalignment}, on the other hand, as found for the HD17156 \cite{Narita08}, XO-3 \cite{Hebrard08} and WASP-14 \cite{johnson09} systems, can be produced by planet scattering or Kozai mechanism due to the presence of a third body (e.g. \cite{Chatterjee08}).
Interestingly, this spin-orbit alignment shows that the star's axis of rotation and the orbit angular momentum evolution are not significantly altered during the early episodes of angular momentum loss characterized by outflows and disk-star magnetic coupling.
A less frequently addressed tidal parameter is the {\it planetary} obliquity, i.e. the  angle between the planetary spin axis and the orbit normal. Although the planet's obliquity is expected to be rapidly damped by tidal dissipation, a persistent nonzero planet
obliquity has been suggested as a result of the capture in
a Cassini 2 spin-orbit resonance state as the nebula dissipates \cite{WinnHolman05}. This scenario, however, has been excluded for hot-Jupiter-like planets \cite{Levrard_etal07,Fabrycky_etal07,Peale08}. Indeed,
the resonant equilibrium is eventually destroyed by the strong tidal torque, leading the system
to leave the Cassini 2 state and spiral towards the Cassini 1 state, with negligible obliquity
 \cite{Levrard_etal07,Peale08}. Therefore, close-in planets quickly evolve to a state with the planet's spin axis nearly normal to the orbit plane, i.e. a negligibly small planet obliquity.
Note that the determination of stellar companion's obliquities
would provide an interesting diagnostic to distinguish planets from brown dwarf companions, these latter being formed from the same original gravoturbulent collapse of the parent cloud as the star, leading to arbitrary spin angle distributions.

\subsection{Orbital evolution}
\label{orbit}
For close planets,
gravitational interactions between the planet and the star lead each of these bodies to raise tides on the other one, generating torques in the tidal bulges. Through exchange of angular momentum,
these strong tidal interactions affect the system orbital (eccentricity, semi-major axis, obliquity) and rotational (stellar and planetary spin, $\omega_\star$, $\omega_p$) properties.
In the first calculations addressing tidal effects in the discovered extrasolar close-in star-planet systems \cite{bodenheimer01,Rasio96}, coplanarization of the orbit and the stellar equator (i.e. stellar spin-orbit alignment), planet spin-orbit alignment,
circularization of the orbit and synchronization of the stellar and planetary rotation with the orbital motion 
were implicitly assumed to represent the asymptotic
{\it equilibrium states}, characteristic of the endpoint of tidal evolution, although the orbit might be {\it unstable}, leading eventually to star-planet merging \cite{Rasio96}. Accordingly, the associated
timescales were generally characterized by an exponential relaxation towards an equilibrium state
\cite{Zahn77,Hut81}.
Furthermore, in the calculations of \cite{bodenheimer01}, only tides raised by the star on the planet were considered whereas tides raised by the planet on the star were ignored. 
A proper derivation of the evolution of all orbital and rotational parameters, however, requires to solve consistently the {\it complete}
non-linear
coupled tidal equations, taking {\it both the planetary and the stellar tides} into account. Note that tides raised in the star transfer angular momentum between the star's rotation and the planet's orbit, so that, in the presence of stellar tides, the planet's orbital angular momentum is {\it not} conserved during the tidal evolution, even if the planet's rotation is synchronized. 
Such consistent calculations lead to markedly different dynamical evolutions with significantly different rotational and orbital evolution timescales, and thus different eccentricity values from the ones obtained when neglecting stellar tides \cite{DobbsDixon04,Patzold04,Jackson08a,Levrard_etal09,IbguiBurrows09,Miller_etal09}.

As mentioned above, most of the early studies implicitly assumed the existence of a tidal (possibly unstable)
minimum energy equilibrium state at given angular momentum for the orbiting planets, characterized by circular orbits (for the stable state), synchronous rotation and spin-orbit alignment. However, as demonstrated recently \cite{Levrard_etal09}, none but one, namely Hat-P-2b, of the transiting planets discovered at this time (26 objects) has such an equilibrium state. Indeed, for all these systems,
the total star-planet angular momentum, $L_{tot}$, is lower than the critical angular momentum defined as
$L_{\mathrm{c}}=4 \left[\frac{G^2}{27} \frac{M^{3}_{\star} 
M_p^{3}}{M_{\star}+M_p}(I_p+I_{\star}) \right]^{1/4}$, where $I_p$ and $I_{\star}$ denote the polar moments of inertia  
of the planet and the star, respectively. Consequently, no equilibrium state exists and all these planets
 will ultimately merge with the star \cite{Cou73,Hut80}.
This result does {\it not depend on any particular tidal model} and bears major consequences on
our analysis of the discovered transiting systems and, possibly, of the non-transiting ones too. First of all,
it implies that the conventional  tidal exponential damping estimates for the timescales for semi-major axis evolution, synchronization of the spins and spin-orbit alignment are not valid, because the implied corresponding
equilibrium states do not represent the endpoint of the tidal evolution. The full tidal calculations, taking into account the strong nonlinear coupling between $e$, $a$, $\epsilon$, $\omega_\star$ and $\omega_p$ show that
the timescales characteristic of the semi-major axis and the stellar spin and obliquity
evolution of these systems are now comparable, and equal
to the lifetime of the system itself \cite{Levrard_etal09}. These quantities are found to evolve only moderately from their
initial values, until they quickly go to zero or diverge during the final merging episode with the host star, the true endpoint of the tidal evolution for these systems. 
The only exception is the
 {\it pseudo}-synchronization of the {\it planet's} rotational velocity with the orbital motion, such as $\omega_p\sim n$, which occurs on a timescale comparable to the one
estimated with the equilibrium tide theory, $\sim 10^5$-$10^6$ yr for hot-Jupiter typical parameters \cite{bodenheimer01,Rasio96,Levrard_etal09}:

\begin{eqnarray}
\tau_{sync,p} =\frac{I_p}{\Gamma_p} |\omega_p - n| \approx 5.0\, (\frac{\alpha}{0.25}) (\frac{Q^\prime_p}{10^6}) (\frac{M_p}{\mjup}) (\frac{R_p}{\rjup})^{-3} \,{P}_d^3\,\,\,{\rm Myr},
\label{tsync}
\end{eqnarray}
where $I_p=\alpha M_p R_p^2$ is the planet's moment of inertia ($\alpha \approx 0.25$ for Jupiter), $\Gamma_p=\frac{3}{2} \frac{GM_\star^2R_p^5}{Q^\prime_pa^6}$ the amplitude of the torque exerted by the star on the planet,
$P_d$ the orbital period in days and $Q^\prime=Q/k$, where $Q_p$ (resp. $Q_\star$) denotes the planet (resp. star) tidal quality factor.
This latter is defined as the ratio of the maximum potential energy stored in the tidal distortion
over the energy dissipated per tidal period, roughly equal to the inverse of the tidal phase lag associated to the (equilibrium) tidal oscillation in the body of interest (e.g. \cite{GoldreichSoter66}), and $k$ denotes the star or planet second order Love number, representative of the body's deformability (typical values of the Love number for Sun-like stars and for Jupiter are $k_\odot \approx 0.02$ and $k_{\rm Jup}\approx 0.38$, respectively, although with large uncertainties).
This quick planet pseudo-synchronisation stems from the fact that
the planet's rotation angular momentum is much smaller than the orbital angular momentum. This state,
however, corresponds to a {\it temporary state}, as the ultimate planet-star orbital collapse eventually
causes the planet to spin up dramatically \cite{Levrard_etal09}. Once the planet is pseudo-synchronized, 
there is no further exchange of angular momentum between its rotation and its orbit; the only exchange occurs between the orbit, the
eccentricity and the stellar rotation at constant total orbital angular momentum. These quantities are also modified by the
tides raised by the star on the planet, in case of a non-zero eccentricity, and by the tides raised by the planet on the star
even when/if the orbit is circularized. 

This lack of tidal equilibrium state for the transiting systems bears other important consequences. First of all, it
implies that, even for a circular orbit ($e=0$), ongoing tidal dissipation keeps taking place in the {\it star}, leading, by conservation of angular momentum, to the planet's endless orbital decay and, inevitably, to a merging of the planet with the star. It also means that {\it stellar} spin synchronization with the orbit can never occur for these systems, since, once the planet is pseudo-synchronized, the tidal torque raised by the planet onto the star yields this latter to {\it spin up constantly} (indeed, for $n\gg \omega_\star$, energy is transferred from the orbit to the spin \cite{Hut81}), the stellar spin evolving within the same timescale as the orbital distance. Decrease of stellar rotation velocity with time due either to stellar winds or
magnetic braking may counteract this tidal spin-up. Strong stellar wind episodes, however, mainly occur at the early stages of evolution and these effects are probably unimportant at the age of the observed systems. Magnetic braking is a more complex issue and is found to compensate or even dominate tidal spin-up if $Q^\prime_\star \gtrsim 10^8$ for OGLE-TR-56b, the presently discovered extrasolar giant planet closest to its parent star \cite{CaronePatzold}.

As mentioned above, a proper determination of the various timescales, including the lifetime of the system, i.e. the time for the planet to merge with its host star (or more exactly to reach the Roche Lobe limit, $a\sim 2.5\,(\rho_\star/\rho_p)^{1/3}R_\star$), requires to solve consistently the complete non linear coupled
dynamical equations for $a$, $e$, $\epsilon$, $\omega_p$ and $\omega_\star$. No analytical solution exists.
In the limit of small eccentricity ($e \ll 1$) and for $M_p\ll M_\star$, 
however, the timescale for orbital evolution is dominated by the tides in the {\it star} \cite{Patzold04,Jackson08a,Levrard_etal09}\footnote[1]{As mentioned in \cite{Levrard_etal09}, even though only the outermost layers of solar-type stars are convective and dissipate efficiently, ultimately the convective and radiative zones will synchronize and dissipation will occur throughout the entire star. This will affect the timescale for tidal evolution, but this latter will qualitatively remain the same.}. In that case,
providing that the orbital period is much shorter than the stellar rotation period ($\omega_\star \ll \omega_{orb}=2\pi/P $), a fulfilled condition for most of the observed transits ($P_{rot,\star} \sim$ 3-70 days whereas $P \lesssim$ 4 days (see e.g. Table I of \cite{Levrard_etal09})), and that the stellar obliquity is small (also a generally fulfilled condition), the orbit evolution timescale, given by the inverse ratio of the torque exerted by the {\it planet} on the {\it star}
over the orbital angular momentum, can be estimated as
(e.g. \cite{GoldreichSoter66,Hut81,Peale99,Patzold04,Levrard_etal09})

\begin{eqnarray}
\tau_a \approx( {\Gamma_\star \over M_p a^2 n})^{-1} 
& \approx & 0.06\, (\frac{Q^\prime_\star}{10^6}) (\frac{M_\star}{M_p}) (\frac{a/0.02\,{\rm AU}}{R_\star/R_\odot})^5 \,{P}_d\,\,\,{\rm Myr} \nonumber \\
& \approx & 0.05\, (\frac{Q^\prime_\star}{10^6}) (\frac{M_\star}{M_p}) (\frac{{\bar \rho}_\star}{{\bar \rho}_\odot})^{5/3}\, {P}_d^{13/3}\,\,\,{\rm Myr},
\label{timing_a}
\end{eqnarray}
Note that this timescale quickly decreases with finite eccentricity since tidal effects increase drastically at the apside \cite{Levrard_etal09}.
We stress again that eqn.(\ref{timing_a}) is only a rough estimate and that the proper determination of the orbit evolution requires a consistent solution of the full tidal equations. Moreover, the true calculation of the timescale depends to some extent on the used tidal model. Eqn.(\ref{timing_a}), however, provides some useful information. Since indeed, at very small eccentricity, $\tau_a$ only depends on (or more exactly is dominated by) tidal dissipation within the star, not within the planet, as seen from eqn.(\ref{timing_a}), a first consequence of this equation is that, for hot-Jupiter like systems:

\begin{equation}
\tau_a \lesssim \langle \tau_\star \rangle \sim 10\,\,{\rm Gyr}\Rightarrow P\lesssim 3.4\,(\frac{Q^\prime_\star}{10^6})^{-3/13}\,\,\,{\rm day},
\label{timing_c}
\end{equation}
where $\langle \tau_\star \rangle$ is the average value of the lifetime of stars harboring planets ($\sim 10$ Gyr for solar-type stars).
Therefore, although all the planets for which $L_{tot}<L_c$ will merge with their host star, for the ones with orbital periods in the
range $P\sim 1$-10 days, i.e. semi-major axis $a\sim 0.02$-0.09 AU, for a range of uncertainty $10^9\ge Q^\prime_\star \ge 10^4$, the merging will occur
within the stellar lifetime. Note from eqn.(\ref{timing_a}) that this planet's lifetime decreases with increasing planet-star mass ratio, which has two major consequences. First of all, the more massive the planet, the faster the orbital decay and thus the shorter the system's lifetime, which may partly explain the lack of hot-Jupiters for $P\lesssim 3$ d \cite{Patzold04,Gu03,Jackson09a}. Second of all, for a given orbital distance
and comparable stellar masses, small transiting planets should be discovered around older systems, a trend which seems to be supported by observations \cite{Jackson09a,Jackson09b}. The often used assumption that tidal evolution is always dominated
by tides raised by the star in the planet leads to the opposite behaviour.
Conversely, as seen from eqn.(\ref{timing_a}), the determination of the age of the system from the age of the host star provides a constraint on the minimum value for $Q_\star$.
As mentioned earlier, for observed transiting planets, the orbital period is shorter than the stellar rotation period, so that tidal dissipation in the central body (the star in the present context) {\it decreases} the orbital semi-major axis (${\dot a}<0$), by conservation of total angular momentum. This is similar for instance to the Mars-Phobos system but contrasts with the Earth-Moon one. In this latter case, the Moon revolution period is longer than the Earth rotation one so that tidal dissipation in the Earth leads to an increase of the Earth-Moon distance.

As for the orbital evolution timescale, there is no simple analytical solution for the orbit circularization timescale, $\tau_e$, and one needs to solve the whole dynamical equation system. One can try, however, to estimate $\tau_e$, as done above for $\tau_a$. The rate of eccentricity decay reads
\begin{equation}
\frac{de/dt}{e}=-(\frac{1}{\tau_{e,p}}+\frac{1}{\tau_{e,\star}}),
\end{equation}
where $\tau_{e,p}$ and $\tau_{e,\star}$ denote the characteristic timescales associated with the contributions steming from the tides raised by the star in the planet and by the planet in the star, respectively. It can easily be shown (e.g. \cite{Goldreich63,Peale99}) that (for zero stellar obliquity)
\begin{equation}
\frac{\tau_{e,p}}{\tau_{e,\star}}=(\frac{M_p}{M_\star})^2 (\frac{R_\star}{R_p})^5 ({Q^\prime_p \over Q^\prime_\star}).
\end{equation}
It can be verified that, for the majority of the presently observed transit planets, $\frac{\tau_{e,p}}{\tau_{e,\star}}<1$ (for $Q^\prime_p \le Q^\prime_\star$), so that the eccentricity damping is dominated by tidal dissipation of the torque exerted by the {\it star} on the {\it planet}. 
In that case, the orbit circularization timescale can be estimated as:

\begin{eqnarray}
\tau_e \approx \tau_{e,p} \approx ({\Gamma_p \over M_p a^2 n})^{-1} 
& \approx & (5.3\times 10^3)\, (\frac{Q^\prime_p}{10^6}) (\frac{M_p}{M_\star})  (\frac{a/0.02\,{\rm AU}}{R_p/\rjup})^5\, {P}_d\,\,\,{\rm Myr} \nonumber \\
& \approx & 5\, (\frac{Q^\prime_p}{10^6}) (\frac{M_\star}{\msun})^{2/3} (\frac{M_p}{\mjup})(\frac{R_p}{\rjup})^{-5}\,{P}_d^{13/3}\,\,\,{\rm Myr}\nonumber \\
& \approx & 0.04\,(\frac{Q^\prime_p}{10^6}) (\frac{M_\star}{M_p})^{2/3}  (\frac{{\bar \rho}_p}{{\bar \rho}_{\rm Jup}})^{5/3} \,{P}_d^{13/3}\,\,\,{\rm Myr},
\label{timing_b}
\end{eqnarray}
As seen from eqns.(\ref{timing_a}) and (\ref{timing_b}) (and as expected from the fact that $\tau_{e,\star} \sim \tau_{a}$ for $e\ll 1$), the ratio of the orbital and circularization timescales in the limit of small eccentricity reads:

\begin{equation}
{\tau_e \over \tau_a} \approx ({M_p \over M_\star})^2  ({R_\star \over R_p})^5  ({Q^\prime_p \over Q^\prime_\star}),
\label{timing_c}
\end{equation}
i.e. ${\tau_e / \tau_a} \approx 0.1\, (Q^\prime_p / Q^\prime_\star)$ for typical hot-Jupiter conditions. Therefore, for $Q^\prime_p / Q^\prime_\star \lesssim 10$, tidal damping may explain the circularization of initially eccentric orbits. Indeed, calculations solving the complete tidal dynamical equations show that tidal dissipation for short-period planets may have produced the present eccentricities from larger
initial values similar to the longer period planet eccentricity distribution, providing an explanation for the smaller eccentricity ($e \lesssim 0.3$, $\langle e \rangle \sim 0.1$) of short-period ($a\lesssim 0.2$ AU) planets compared with more remote ones ($e \lesssim 0.9$, $\langle e \rangle \sim 0.3$), depending on the values of $Q_\star^\prime$ and $Q_p^\prime$
 \cite{Jackson08a,Jackson08b,IbguiBurrows09,Miller_etal09}. 
For $Q^\prime_p / Q^\prime_\star \gtrsim 10$, however, the orbit does not have time to circularize within the lifetime of the system \cite{Matsumura08,Levrard_etal09}. In such cases, there is no need to
invoke undetected companions to maintain the eccentricity. Note that a value $e$=$0$ is often assumed when constraining the orbital parameters by inferring a solution for $(a,e$) from the observed light curves in fitting the orbits of close-in planets. Eccentricity values for short-period planets should thus be taken with caution.

Integrating the {\it complete} coupled tidal evolution equations back in time from present eccentricity and orbital values\footnote{Present attempts to address this issue \cite{Jackson08a,IbguiBurrows09,Miller_etal09} do not consider the evolution of the stellar spin and use tidal equations truncated at the order $e^2$, only valid for small eccentricity values, and thus yield quantitatively unreliable results.} may provide some clues about
the initial eccentricity distribution of exoplanets, at the end of their formation process, and thus about the outcome of planet-disk interactions. Numerous Lindblad or corotation resonances take place between a disk and a planet,
exerting torques which deposit or remove angular momentum from the resonance locations. This modifies the planet's eccentricity during the disk evolution \cite{GoldreichSari03}. The net effect of these
disk-planet interactions on the eccentricity, however, remains uncertain, and it is not clear whether
eccentricity is damped or excited, with possibly both solutions being valid depending on the disk and planet properties \cite{MooreheadAdams}. To first order, however, the eccentricity evolution due to disk torques is ${\dot e}\propto 1/\sqrt{a}$ \cite{GoldreichSari03}, so it is expected that short period planets are
more affected than the ones located further away. Planets could thus form with finite eccentricities, damped either by interactions with the disk \cite{MooreheadAdams}
or by tidal interactions after the disk dissipation for the low eccentricity objects. On the other hand, large initial eccentricities can be produced by planet-planet scattering or gravitational perturbations by a stellar companion \cite{FordRasio08,Chatterjee08}.

As discussed in the next subsection, the orbital and rotational evolution timescales, however, can not be determined accurately, as they depend on
the ill-determined tidal forcing and dissipation processes, as well as on the internal structure of the star and the planet (radiative envelope, presence of a dense core, etc...), which translates into values of $Q$ uncertain by several orders of magnitude.

\subsection{Tidal energy dissipation}
\label{tidal_dissipation}

If the planet's orbit is circular, the tidal bulge is motionless in the planet's frame and there is no energy release. In the case of a maintained finite eccentricity or obliquity, however, the persisting tidal friction
leads to energy dissipation in the planet, at the expense of the orbital and rotational energies. 
 If the tidal heating induced in the planet by tidal dissipation dominates the planet's main sources of energy, gravothermal contraction and solar irradiation, about
$10^{25}$ erg s$^{-1}$ for typical Gyr-old hot-Jupiters ({\it i.e.} $\sim 1\, \mjup$ at $\sim 0.05$ AU from a Sun-like star) and lasts over a timescale comparable to the planet's thermal timescale, this
extra source of energy will slow down the planet's contraction, leading to a larger radius (and luminosity) at a given age than otherwise expected. 
At second order in eccentricity (i.e. in the limit $e\ll 1$), the tidal dissipation rate in a pseudo-synchronously rotating ($\omega_p\simeq n$)
planet with zero obliquity is given, at given $a$ and $e$, by (e.g. \cite{PealeCassen78,Hut81,Levrard08,Ferraz-Mello08}):

\begin{eqnarray}
\frac{dE_{tide}}{dt} &=&  - \frac{21}{2} \frac{G^{3/2}M_\star^{5/2}} {Q_p^\prime} (\frac{R_p^5}{a^{15/2}}) \,e^2 \nonumber \\
&\approx& -5.2\times 10^{27}\,\,( \frac{Q_p^\prime}{10^6})^{-1}\,( \frac{M_\star}{M_\odot})^{5/2}\, {(R_p/R_{\rm Jup})^{5}\over (a/0.05\,{\rm AU})^{15/2}} \,e^2 \,\,\,\,{\rm  erg\, s}^{-1},
%
\label{Edot}
\end{eqnarray}
where the terms of orbital evolution only include the contributions due to the planetary tides, since we are presently interested in the energy dissipated in the planet.
Tidal dissipation has been proposed to explain the abnormally large radii of some transiting planets \cite{bodenheimer01}.
As mentioned earlier, however, these estimates of orbital decay timescales were 
ignoring the stellar tides, leading to incorrect tidal dissipation rates. As a consequence, eccentricity and tidal dissipation were found to be negligible at the age of the presently observed transiting planets, and hypothesis like continuous gravitational interactions from undetected companions had to be invoked to maintain a finite eccentricity. 
Only recently have correct tidal evolution calculations been conducted to quantify this effect \cite{DobbsDixon04,Patzold04,Jackson08a,Jackson08b,Levrard_etal09,IbguiBurrows09,Miller_etal09}. As mentioned earlier, the coupled tidal
equations taking both star and planet tides into account yield significantly different timescales
for the evolution of the orbital and rotational properties, with a non-monotonic evolution for the evolution of the eccentricity, semi-major axis and stellar spin,
as well as for the tidal heating, and
this latter can provide in some cases
a significant enough source of energy to alter the cooling properties of the planet at a few Gyrs. 

For tidal energy dissipation to significantly affect the contraction and thus the radius of the planet, however, tidal heat must be deposited deep in the convective layers.
Turbulent viscosity (characteristic of what is usually refered to as "equilibrium tide") may lead to dissipation of heat at depth \cite{Zahn66}. However, for the planets of interest, the convection turnover time exceeds the tidal period, so that convective dissipation is probably too slow to be efficient in short-period planets as well as in their host stars \cite{GoldreichNicholson77,Zahn89}. Excited short wavelength $g$-modes (characteristic of "dynamical tides") at the interface between convective and radiative zones \cite{Zahn70} would dissipate within the radiative region, since $g$-modes do not propagate in isentropic regions, only affecting the outermost parts of irradiated planets. An interesting alternative mechanism in rapidly rotating planets might be short-wavelength, low-frequency inertial waves, restored by Coriolis rather than buoyancy forces, excited at the envelope-core boundary in the central parts of the planet. These waves can propagate and might dissipate in the convective inner regions and are thus more prone to affect the planet's structure than the aforementioned gravity modes \cite{OgilvieLin04,Wu05,GoodmanLackner09}. In that case, the tidal quality factor is found to depend on the orbital period and core size as $Q^\prime \propto R_c^{-5}P^2$ \cite{GoodmanLackner09}.
 As discussed above after eqn.(\ref{timing_a}), the inferred age of the systems implies that $Q_\star^\prime$ can not
be substantially smaller than the nominal values $Q_\star^\prime\approx 10^5$-$10^6$. The same holds true for $Q_p^\prime$.
Indeed, most of the transiting planets are predicted to have a core whose state remains uncertain (see \S \ref{eosz}). A solid
core implies a much lower quality factor, $Q_{core}^\prime\sim 10^2$-$10^3$, than for a fluid or gaseous body,  i.e. a more dissipative planetary interior, which in turn increases
the expected tidal dissipation ($dE_{tide}/dt\propto 1/Q_p^\prime$). The
eccentricity of Io's orbit, however, implies that $6\times 10^4\lesssim Q_{\rm Jup}\lesssim 2\times 10^6$ \cite{YoderPeale81}. Moreover, as mentioned above, if inertial waves are the main culprit for tidal energy dissipation, $Q_p^\prime$ is predicted to be of the order of $\sim 10^6$ for Jupiter (Io's orbital period is $P=1.77$ d) and $\sim 10^7$-$10^8$ for hot Jupiters ($P\sim 3$-4 d). This suggests that the jovian planet cores may have melted, under the action of
the violent accretion episodes during the formation or of tidal heating during the evolution. 

Besides the gravitational tides mentioned in this section, it is worth mentioning the possibility for short-period, strongly irradiated planets to experience {\it thermal tides} \cite{Arras09}. Thermal tides arise from the torque exerted by the stellar gravitational field on the thermal bulge produced in the planet's outer layers by the stellar time-dependent insolation. In that case, an equilibrium state in which gravitational and thermal tide effects balance each other can be reached, leading to asynchronous spin and/or possibly to finite eccentricity. This process, however, has been shown not to
be valid in the case of gaseous bodies since, in the absence of a surface crust, the torque exerted on the thermal bulge vanishes \cite{Goodman09}.

\section{\label{constraints} Observational constraints}

\subsection{\label{section_transit} The radius anomaly of transiting exoplanets}

As mentioned in \S \ref{section_observation} a significant fraction of transiting exoplanets exhibits abnormally large radii (see Fig. \ref{transit_mr}).  This puzzling property is one of the most titillating observational discoveries in this new field and still requires a robust explanation. Several ideas have been proposed, as described below, but none of them
has yet received a consensus.

\subsubsection{Atmospheric circulation}
\label{atm-circ}

As mentioned in \S \ref{section_cooling}, close-in planets receive a substantial
amount of energy from the star. The resulting heating of the outer
planet layers affects the vertical temperature stratification and, consequently, the planet's evolution and thus radius at a given age,
as described in \S \ref{section_cooling}.
 This effect, however,  is insufficient to explain  observed radii 
 $\simgr 1.2\, \rjup$. Irradiation can also modify the horizontal temperature profile. Indeed, if the planet is tidally locked 
and receives constantly the stellar flux on the same hemisphere, strong temperature contrasts can exist  between the day- and night-sides of the planet, inducing 
 rapid atmospheric circulation (see \S \ref{section_atmosphere}). The resulting fast winds may produce a heating mechanism
in the deep interior of the planet, slowing down its evolution and yielding a larger than expected radius. This suggestion was based on
 numerical simulations of atmospheric circulation  by \cite{showman02}, which produce a downward flux of kinetic energy of about $\sim$ 1\% of the absorbed stellar flux. This
 heat flux is supposed to dissipate in the deep interior, producing an extra source of energy during the planet's evolution. According to these simulations, this mechanism yields a typical heating rate of 10$^{27}$ erg s$^{-1}$,
 which can explain the inflated radius of hot Jupiters like HD209458b (\cite{guillot02,chabrier04}.
 The validity of this scenario, however, is still debated. The substantial transport of kinetic energy found in the simulations of \cite{showman02} strongly depends on the details of atmospheric circulation models and has not been confirmed by other
 simulations (\cite{burkert05}). Although important efforts are devoted to the 
 development of
 multi-D dynamical simulations of strongly  irradiated atmospheres (see \S \ref{section_atmosphere}), the various simulations so far yield very different results (\cite{showman07}). Furthermore, as demonstrated by \cite{goodman09b}, a proper description of the physical mechanisms responsible for energy dissipation and drag in the flow ({\it e.g.}, shocks, turbulence) is mandatory to obtain reliable descriptions of heat redistribution and wind speeds across irradiated exoplanetary atmospheres.
  
\begin{figure}
\begin{center}
\includegraphics[width=10cm]{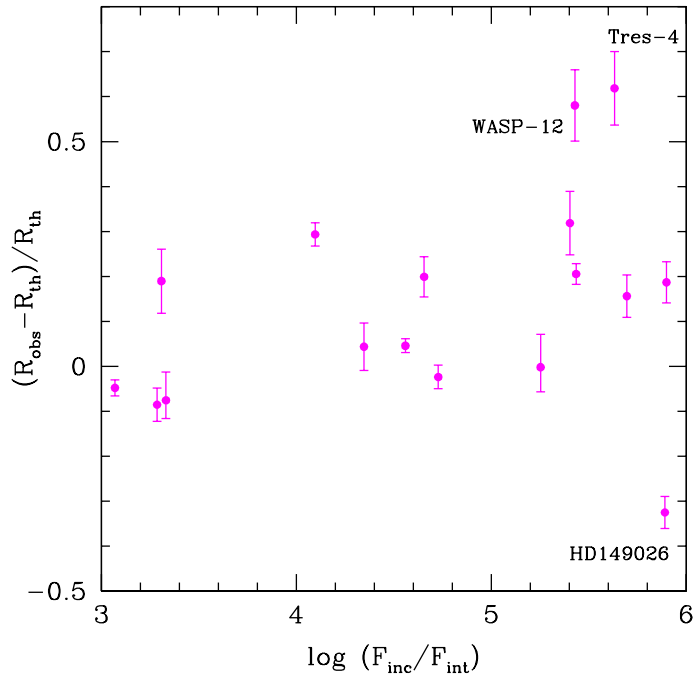}
\end{center}
\caption{Radius excess of transiting exoplanets as a function of the incident stellar flux 
$F_{\rm inc}$ normalised to the planet's predicted intrinsic flux
at the age of the system, $F_{\rm int}={L_{\rm int} \over 4 \pi R_{\rm th}^2}$. 
The radius excess is defined as ${ R_{\rm obs} - R_{\rm th} \over
R_{\rm th} }$, where  $R_{\rm obs}$ is the observed radius and $R_{\rm th}$
is the predicted radius of the planet at the age of the system. The error
bars on the radius correspond to the quoted observational values.} 
\label{finc}
\end{figure}

Assuming this mechanism is at play in all close-in planets, and arbitrarily adding an extra-source of energy corresponding to 0.5\% $\times \finc$, \cite{guillot06} find a correlation between the mass of heavy elements required to fit the transit radii and the metallicity of the parent star. This result, however, is highly speculative, since it crucially depends on the aforementioned assumption of a {\it constant} fraction of $\finc$.
If this wind-induced mechanism is at play in all close-in planets,
one expects a correlation between the planet radius "excess", defined as the relative difference between the observed radius $R_{\rm obs}$  and the theoretical radius $R_{\rm th}$ obtained with regular irradiated models, ${ R_{\rm obs} - R_{\rm th} \over
R_{\rm th} }$, and the incident stellar flux $\finc$, normalised to the planet's intrinsic flux $F_{\rm int}={L_{\rm int} \over 4 \pi R_{\rm th}^2}$, with $L_{\rm int} = \int -T {dS \over dt} \, dm$, where $S$ is the specific entropy of the irradiated planet.
This relation is illustrated in Fig. \ref{finc} for a handful of transiting systems. 
Whether the expected trend exists or not is not clear, in particular when considering the fact that WASP-12's large radius may be explained by tidal heating, as mentioned in \S \ref{tidal} below. Clearly, a more detailed analysis, extended to all transiting planets, is required to explore this issue.

To summarize, although atmospheric circulation remains an attractive possibility to explain the "hot-Jupiter" radius anomaly,
a robust confirmation of this mechanism, in particular of heat transport at deep enough levels to affect the internal adiabat,
is needed, based on more sophisticated numerical simulations. It should also be noted that planets far enough from their host star ($\simgr 0.1$ AU) should
not be affected by this mechanism, a diagnostic eventually testable with the Kepler
mission.

\subsubsection{Tidal effects} 
\label{tidal}

\begin{figure}
\begin{center}
\includegraphics[width=10cm]{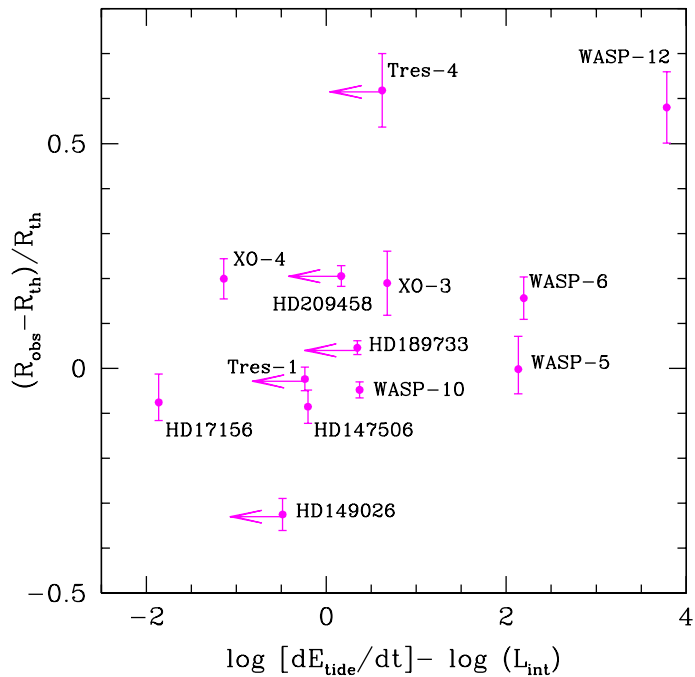}
\end{center}
\caption{Radius excess  (same definition as in Fig. \ref{finc}) of transiting exoplanets as a function of the tidal dissipation contribution (see eqn.(\ref{Edot})), assuming constant values for $a$ and $e$,
 at the age of the system. A value $Q^{'}_p=10^6$ is assumed for the planet's tidal quality factor. Objects shown  
with a left arrow  have a (measured or assumed) eccentricity $e \le 0.01$, and their tidal dissipation is calculated here assuming $e=0.01$. }
\label{etide}
\end{figure}

Following \cite{bodenheimer01}, several studies have suggested 
tidal heating  as a possible mechanism to explain
inflated transiting planets (see \S \ref{tidal_dissipation}). 
A heating rate of the order of 10$^{27}$ erg s$^{-1}$, 
consistent with the tidal dissipation rate given in eqn.(\ref{Edot}) 
and about 100 times the typical intrinsic energy flux of hot-Jupiters, 
dissipated in the planet's convective interior over appropriate timescales could in principle explain the observed abnormally 
large radii for at least some {\it coreless} planets. 
The presence of a central core yields a smaller radius and thus requires 
a larger energy rate. As discussed in \S \ref{tidal_dissipation},
the first calculations (\cite{bodenheimer01}) exploring the effect of tidal dissipation
were based on incorrect tidal evolution timescales and assumed constant orbital properties over time. 
The correct impact of tidal dissipation on the planet radii from their supposed initial semi-major axis and eccentricity, taking into
account the opposite effects of decreasing $a$ and $e$ on the tidal dissipation rate $\dot E$ along the evolution (see eqn.(\ref{Edot})) has been explored only recently \cite{Jackson08b,IbguiBurrows09,Miller_etal09}, although under some restrictive conditions (see second footnote in \S \ref{orbit}). In some cases, this heat source is found to be substantial at the age of the system, possibly affecting the planet's contraction and helping solving the radius
discrepancies, {\it assuming tidal dissipation occurs deep in the convective interiors} (see discussion in \S 7.3). 
As shown in Fig. \ref{etide}, WASP-12 experiences extreme tidal heating, which may explain its
strikingly large radius (see \cite{Miller_etal09}). 
Tidal heating due to finite eccentricity, however, does not seem to be the lacking dominant mechanism
responsible for {\it all} the observed abnormally large radii. Indeed, as illustrated in 
Fig.\ref{etide} (see also \cite{IbguiBurrows09,Miller_etal09}), there is no obvious
correlation between abnormally inflated planets and tidal dissipation energy.
Tres-4, for instance, experiences a tidal heating comparable to its internal energy rate, even though being notably inflated (see \cite{Miller_etal09}). However, proper calculations of the complete (not truncated at a given eccentricity order and including the rotational energy) tidal evolution equations remain to be done to fully address this issue.

An other source of tidal heating can arise if the orbit is inclined relative to the stellar quadrupole, so that
the planet makes a vertical oscillation in the quadrupole field of the star once per orbit. This effect, however, has been shown to be utterly negligible \cite{Winn_etal05}. Finally, as mentioned in \S \ref{tidal_dissipation}, thermal tides can not be maintained in gaseous planets and thus can not provide an extra source of heat \cite{Goodman09}.


\subsubsection{Atmospheric enhanced opacities} 

A possibility to slow down the contraction of a self-gravitating body and to obtain a larger radius at a given age is to increase the atmospheric opacities. This solution 
was recently suggested by \cite{burrows07} to explain the puzzling large 
radii of some transiting planets. 
Alteration of atmospheric properties (chemistry, clouds, opacities) by 
strong optical and UV irradiation or enhanced atmospheric metallicities are suggested as
the source of the increased opacities, although the authors do not  work out the
physical processes that could possibly lead to such an increase. 
In order to mimic the effect of enhanced opacities, they 
assume supersolar atmospheric metallicities (up to  10 times solar). 
The combined effects of enhanced opacities and of the presence of a dense core is found to reproduce the observed 
radius spread of transiting planets (see Fig. \ref{transit_mr}).
Interestingly enough, \cite{burrows07} also finds
the same correlation as suggested by \cite{guillot06}
between planet core mass and stellar metallicity.
Although enhancement of the atmospheric opacities of irradiated extrasolar planets is by no means excluded,
given our limited knowledge of the various physical processes at play in such situations, the results obtained by \cite{burrows07}, based on the
increase of atmospheric metallicities,
remain questionable. First of all, similar consistent calculations by \cite{guillot08b} yield a {\it smaller} radius for the planet, as the larger mean molecular weight due to the increased metallicity counterbalances and even dominates the
slower contraction due to the enhanced opacities.
Second of all, although significant heavy element 
enrichment is observed in the atmosphere of our Solar System
 giant planets (see \S \ref{section_SS}), convective transport in these (cool) planets is
 predicted to occur all the way up to the atmosphere, constantly bringing up heavy material to the atmospheric layers. Irradiated planet atmospheres are radiatively stable down to deep levels. To maintain heavy elements high up in these stably stratified outer layers requires a (yet to be worked out) efficient mechanism to counteract gravitational settling. More importantly, a stratification with heavy elements on top of a (lighter) H/He envelope, as done in \cite{burrows07} has been shown to be unstable, leading to "salt-fingers" instabilities \cite{stevenson85}. 
 The effect of strong irradiation, altering
the chemistry and cloud properties 
requires additional work, as pointed out by the authors themselves \cite{burrows07}.
Therefore, although enhanced atmospheric opacities in strongly irradiated planets is a possibility, it 
requires more robust physical foundations to be considered as the source of the abnormally large radii of irradiated planets.
In any event, even such super-solar metallicity models cannot explain the most inflated transiting planets presently detected, like Tres-4b and WASP-12, implying the need for one or several alternative mechanisms (\cite{burrows07,IbguiBurrows09}).

\subsubsection{Reduced interior heat transport} 

The idea of Stevenson that convection may not be as efficient as usually assumed in planetary interiors 
was resumed and applied to the case of exoplanets by \cite{cb07}. These authors show that conditions prevailing in giant planet interiors could be  favourable to the development of
double-diffusive (layered) or oscillatory convection, if a molecular weight gradient is present (see
 \ref{section_transport}). With characteristic Prandtl number (i.e. the ratio of kinematic viscosity over thermal diffusivity), Pr $\sim$ 10$^{-2}$-1, and inverse Lewis number, defined as the ratio  of solute to thermal diffusivity, Le$^{-1} \sim 10^{-2}$, conditions in planets are not too different from those found on Earth. Indeed, double-diffusive convection, characterised by the formation of multiple thin diffusive layers surrounded by (small-scale) convective regions,
  is a well known process in laboratory experiments, or in some parts of oceans and
  salty lakes, where the stabilizing gradient is due to salinity (the so-called thermohaline convection), with characteristic numbers Pr =7 and Le$^{-1}=10^{-2}$ for salty water \cite{Schmitt94}. 
 The presence of a molecular weight gradient in the interior of giant planets
 could be inherited from the formation process, during the accretion of gas and planetesimals, or due to core erosion. Assuming a molecular weight gradient in the inner parts of a Jupiter-like planet, \cite{cb07}
  show that if multiple layers can form, they might survive long enough to affect substantially the planet evolution. Heat transport  by layered convection is indeed much less efficient than large-scale convection and the heat escape from
the hot, deep parts of the planet is significantly reduced. The upshot is a larger planet at a given age than its homogeneous, adiabatic counterpart (\cite{cb07}). Depending on the number of layers and the steepness of the molecular weight, the presence of this process could explain the observed spread in transit planet radii. If confirmed, this scenario would bear important consequences
on our general understanding of planetary internal structure. High spatial resolution multi-dimensional numerical simulations, which are progressing at a  remarkable pace, might be able in a foreseeable future to confirm or to reject
the existence of layered or oscillatory convection under planetary conditions. Note that, according to this scenario, one should not expect any correlation between the radius and the incident flux, contrarily to other suggested mechanisms like the atmospheric circulation outlined in \S \ref{atm-circ}.
 
\subsection{\label{section_superjupiter} Brown dwarfs versus massive giant planets} 

\begin{figure}
\begin{center}
\includegraphics[width=10cm]{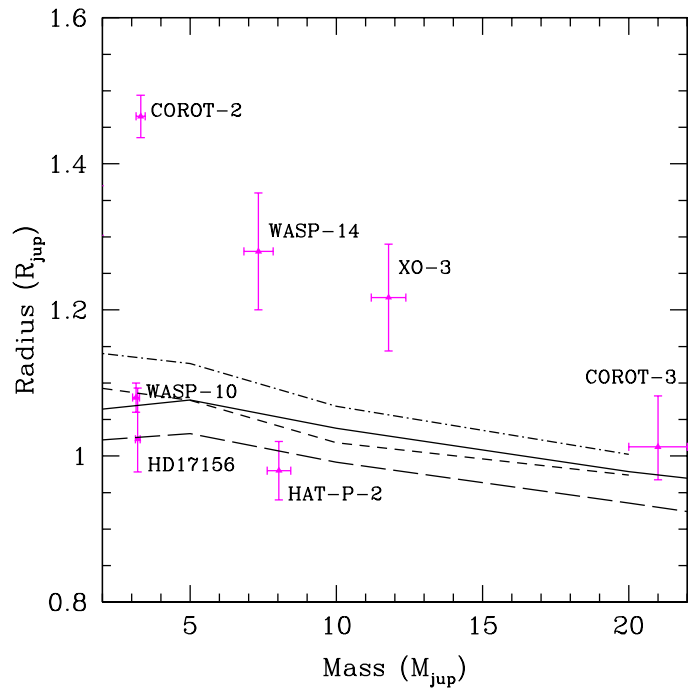}
\end{center}
\caption{Mass-radius relationship for massive transiting exoplanets. The curves
show theoretical predictions at 1 Gyr (models from \cite{baraffe08,leconte09}). Solid line: solar metallicity without irradiation; long-dash: heavy element mass fraction  $Z$=10\% 
and without irradiation; dash-dot: solar metallicity and irradiation from a Sun at 0.05 AU;
short-dash: heavy element mass fraction  $Z$=10\% 
 with irradiation from a Sun at 0.05 AU.
The name of objects in the overlapping mass regime between planets
and brown dwarfs are indicated.}
\label{superjup}
\end{figure}

As underlined  in \S \ref{intro}, brown dwarfs and planets overlap in mass. Observations
of low mass stars and brown dwarfs in young clusters suggest a continuous
mass function down to $\sim$ 6 $\mjup$, indicating that very low mass objects form as an extension of the star formation process \cite{caballero07}. 
There is indeed strong observational support for the brown dwarf formation process to be similar to the stellar one \cite{Luhman07,Duchene09}, and
analytical theories of star formation from the gravoturbulent collapse of a parent cloud do produce proto-brown dwarf cores in adequate numbers compared with the observationally determined mass spectrum \cite{PadoanNordlund04,HennebelleChabrier08, HennebelleChabrier09}. In parallel, the discovery of "super" Jupiter-mass objects $> 10\, \mjup$ orbiting a central star (see Fig. \ref{superjup})
emphasizes the need for identification criteria enabling the distinction between
a brown dwarf and a planet. Radii significantly smaller than
predictions for solar or nearly-solar metallicity objects would reveal the presence of
a significant amount of heavy material, an unambiguous signature of planetary formation process.
The case of the 8 Jupiter mass transiting planet HAT-P-2b (or HD147506b, \cite{bakos07}) is interesting
in this context, as its radius is consistent with  the presence of a significant amount of heavy elements ($\simgr 10\%$ in mass fraction) and cannot be reproduced by a solar metallicity brown dwarf model (\cite{leconte09}), as illustrated in Fig. \ref{superjup}.  If confirmed, HAT-P-2b might be the illustration that planets can form by core accretion up to a least
8 times the mass of Jupiter.
The nature of CoRoT-Exo-3b, a $\sim$ 21 $\mjup$ object orbiting an 
F3V type star (\cite{deleuil08}), however, remains ambiguous, given the present
uncertainties on the radius determination (\cite{leconte09}).
Enhanced metallicity is also expected to leave its mark on the atmospheric properties of a planet and may provide signatures of its formation process in a proto-planetary disk. However, abundance patterns in the cool atmospheres of planets and brown dwarfs  may be severely affected by different complex processes (non-equilibrium chemistry,
cloud formation, photochemistry, see \S \ref{section_atmosphere}). Predictions are thus uncertain and 
 determination of clear spectral diagnostics requires much additional work before they can be used to distinguish a planet from a brown dwarf (see \cite{chabrier07}).

\subsection{\label{neptune} Hot Neptunes and evaporation process} 

About twenty planets with masses less than $\sim$ 20 $\mearth$ have been 
discovered by radial velocity surveys. Since these small planets are at the detection threshold, this number is relatively large, indicating that they should be rather common. Because of their low mass, their presence at close distance from their parent star raises the question about their origin. Observational evidences that close-in planets may undergo significant evaporation process (\cite{vidal03})
gave rise to the idea that hot-Neptunes may have formed from more massive progenitors 
 and have lost a substantial amount of their gaseous envelope (\cite{baraffe05}). This idea and the interpretation, in terms of evaporation, of the observations of \cite{vidal03}  are currently controversial (\cite{benjaffel07}). Preliminary models of evaporation for hot-Jupiters induced
 by the stellar XUV radiation predicted large evaporation rates which could significantly
 affect the planet evolution  (\cite{baraffe04}). More recent theoretical works based on improved treatment of atmospheric escape now reach  different conclusions and find much smaller rates than predicted just after the observations of \cite{vidal03} (see \cite{murray08} and references therein). Small evaporation rates are also consistent with the observed mass function of exoplanets (\cite{hubbard07}). 
 Furthermore,  formation models based on the core-accretion scenario can easily produce hot-Neptunes composed of a large core of heavy material without
 the accumulation of a substantial gaseous envelope (\cite{ida05, alibert06, mordasini09}).
 Their existence can thus be explained without the need to invoke strong evaporation processes.
 Getting the final word requires
 (i) more statistics in the low planetary mass regime, (ii) confirmation of the observations
 and interpretation of  \cite{vidal03} and (iii) improved  models of evaporation of close-in planets. 
 
Independently of these issues, the interior properties of  hot-Neptune planets were
recently revealed by the discovery of  GJ 436b (\cite{gillon07}) and HAT-P-11b (\cite{bakos09}), two transiting planets of $\sim$ 22 $\mearth$ and $
\sim$ 26 $\mearth$, respectively.
 They are remarkably analogous to Uranus and Neptune, in terms of heavy material content, with a radius indicating heavy element enrichment greater than 85\%. For GJ 436b, models with $\sim$ 20 $\mearth$ core of ice or rock and a gaseous H/He envelope of $\sim 2$ $\mearth$ reproduce its observed radius  (\cite{baraffe08}). HAT-P-11b also appears to be a super-Neptune
 planet with $\sim$ 90\% heavy element and $\sim$ 10\% H/He envelope (see Fig. 14
 of \cite{bakos09}).
 Given current uncertainties on the modelling of planetary interior structures, only their bulk composition can be inferred, and the total amount of heavy material, its composition and distribution within the planet cannot be accurately determined. As mentioned in \S \ref{eosz} and shown in \cite{baraffe08}, the thermal contribution of heavy elements to the cooling of Neptune-like planets significantly
affects their evolution and thus radius at a given age. The discovery of GJ436b and HAT-P-11b confirms the large 
heavy element content that can be expected in planets, comforting our general understanding of planet formation through the core accretion model. However, since 
there is no obvious signature of an evaporation process on the structure of a planet,
current observations cannot certify whether or not these hot-Neptunes have undergone strong evaporation episodes during their evolution.

\subsection{Light of extra-solar planets}

\begin{figure}
\begin{center}
\includegraphics[width=11cm]{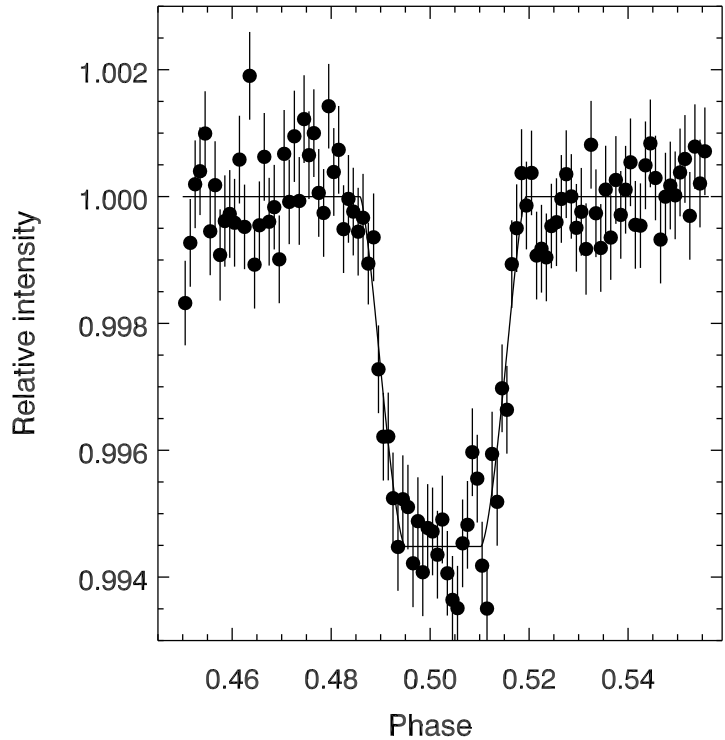}
\end{center}
\caption{Secondary eclipse photometry of HD 189733 at 16 $\mu$m, as measured by
the IRS instrument on board {\em Spitzer} \cite{Deming2006}.}
\label{ecl}
\end{figure}

\begin{figure}
\begin{center}
\includegraphics[width=10cm]{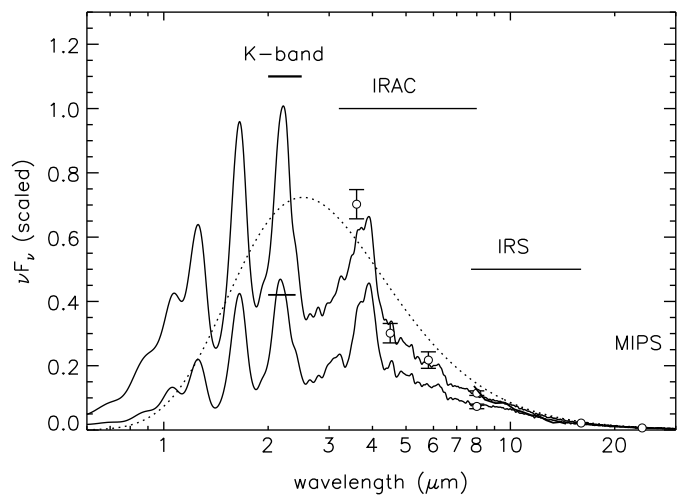}
\end{center}
\caption{HD189733b energy distributions assuming full-redistribution (lower 
line, $\alpha = 0.25$) and no-redistribution (top  line, $\alpha = 0.5$)
along with a 1450K blackbody spectrum (dotted line).  Over plotted are IR
ground and space-based flux measurements for HD189733b \cite{Barnes2007,
Charbonneau2008, Deming2006, Grillmair2007}. The lower point at 8 $\mu$m is
the night side flux measurement from \cite{Knutson2007b}.}
\label{fnu}
\end{figure}

So far, exoplanets have only been discovered by indirect techniques, without
actually direct photon detection from the planet. Various claimed direct detections cannot
unambiguously assess the very nature of the observed object, brown dwarf or planet. The recent detections of the
3-$\mjup$ planet Fomalhaut-b \cite{Kalas2008} and of the young (30 to 160 Myr old) triple system orbiting HR 8799
\cite{Marois2008}, however, represent exciting discoveries and, if coplanarity of the system is clearly demonstrated for the second case, might
indeed be the first genuine direct detections of exoplanets. 
Somewhat counter intuitively, the
short period hot-Jupiters were the first class of exoplanets to have their
atmospheric properties measured, both photometrically and spectroscopically; a
major contribution made possible with the {\em Spitzer Space Telescope}
\cite{Charbonneau2005, Deming2005b}.  The majority of {\em Spitzer} exoplanet
observations target transiting EGPs orbiting nearby stars.  As a transiting
planet disappears behind its star, there is an observable drop in the total
flux from the system as the planet's contribution is temporarily absent.  This
event is often called the secondary eclipse, and the depth of the eclipse is a
direct measure of the planet-star flux ratio (e.g. Fig \ref{ecl}).  The orbital phase
coverage of these Spitzer data confine them to measuring the temperature and
chemistry of a planet's day side only.  Since the planet-star flux ratio is the
quantity measured, secondary eclipses depths increase with increasing
wavelength roughly in accordance with the Rayleigh-Jeans approximation,
resulting in deep eclipse amplitudes even at wavelengths where the planet flux
is quite small.  Figure \ref{fnu} compares the full set of secondary eclipse
flux measurements for HD189733b to dayside model spectra with uniform global
redistribution of absorbed stellar flux ($\alpha = 0.25$) and dayside-only
redistribution ($\alpha = 0.5$).  In this example, there is a transition
between 2 and 3 $\mu$m from one case to the other, suggesting depth-dependent
energy redistribution, probably caused by horizontal flows over a narrow range
of pressures \cite{Barman2008}.

The anticipated diversity in hot-Jupiters is beginning to emerge from secondary
eclipse observations with hints that some hot-jupiters have temperature
inversions across their photospheres (HD209458b and TrES-4) while others may not
(HD189733b).  Temperature inversions can be produced by a variety of processes,
for example, the presence of a high-atmosphere opacity source could produce
strong heating at low pressures.  Plausible opacity sources could be the
molecules TiO and VO which can play an important role in shaping the
atmospheric structures \cite{Hubeny2003, Fortney2008}. However it has yet to be

shown that an identifiable opacity source can produce a temperature inversion
deep (high pressure) enough to reach the photosphere and impact the IR fluxes.
Alternatively, strong horizontal currents could cool deeper layers while
leaving layers above still quite hot resulting in a fairly deep inversion
\cite{Showman2008, Barman2008}.  The exact nature of the inversion and its
cause will require observations of higher precision and across a broader range
of hot-Jupiters.  {\em Spitzer} flux measurements have also started to identify
those planets that have strong or weak day-to-night redistribution of absorbed
stellar energy leading to cold or warm night side temperatures.  Observations
covering a large fraction of an orbital period can provide potential maps of
the atmospheric temperatures allowing tests of global circulation
models \cite{Rauscher2008, Williams2006}.  In one example, HD189733b, the
hottest location of the atmosphere (measured at 8 $\mu$m) is actually not
located at the substellar point but instead leads the sub-stellar point by 2 to
3 hrs (i.e. down wind by 20 to 60$^\circ$) -- as roughly predicted by global
circulation models \cite{Cooper2006}.  For a more comprehensive review of {\em
Spizer} exoplanet observations see \cite{Deming2008}.

Ground-based secondary eclipse observations at K- and L-bands are being
attempted by several groups and, so far, only upper limits have been obtained 
\cite{Richardson2003pt1, Richardson2003pt2, Snellen2005, Barnes2007}. These
near-IR flux upper limits, however, can be extremely useful for estimating the
level of day-to-night redistribution since, in many cases, the peak of the energy
distribution for hot-exoplanets is actually in the near-IR, rather than at
wavelengths covered by {\em Spitzer}.  As demonstrated by \cite{Barman2008} the
question of redistribution is fundamentally a bolometric argument and measuring
the fluxes at multiple band passes including shorter wavelengths is critical for
determining the true day-to-night flux redistribution.

\begin{figure}
\includegraphics[width=16cm]{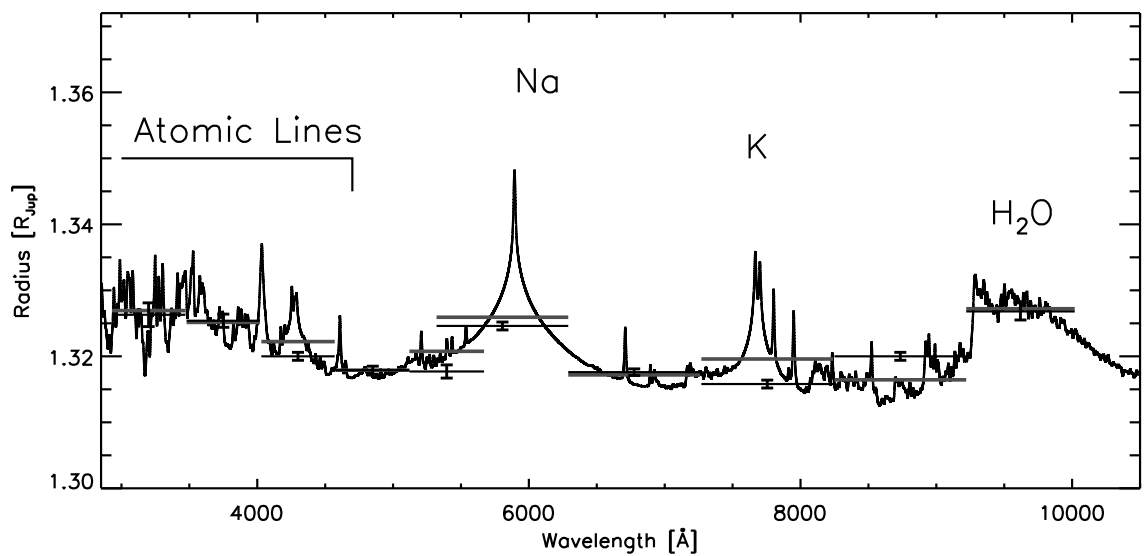}
\caption{Model monochromatic transit radii for HD209458b. Horizontal bars correspond
to mean radii across bins with $\lambda$-ranges indicated by the width of each
bar. Observations are shown with 1-$\sigma$ error-bars
\cite{Knutson2007b, Barman2007}. }
\label{tspec}
\end{figure}

As a transiting planet passes between Earth and the host star, the
wavelength-dependent opacities in the planet's atmosphere obscure stellar
light at different planet radii, leading to a primary transit eclipse with a
wavelength-dependent amplitude.  Spectroscopic observations during primary
transit (called transit spectroscopy) can measure planet radii across various
absorption features from which atomic and molecular mole fractions can be
inferred.  There have been a number of models for EGP transmission spectra
\cite{Seager2000, Brown2001, Hubbard2001, Barman2003, Fortney2003,
Tinetti2007b, Barman2007}, all using 1-D plane-parallel or spherically
symmetric geometry and a single 1-D temperature-pressure profile to represent
the depth-dependent (radial) structure.  The broad Na and K doublets in the
optical were predicted to produce large radius variations \cite{Seager2000}.
Subsequent observations using the STIS instrument on board HST detected the Na
doublet (\cite{charbonneau02}), though it was much weaker than predicted, prompting speculations about high
clouds and photoionization \cite{Barman2003, Fortney2003}.  Following the Na
detection, an extended hydrogen atmosphere surrounding HD209458b was discovered
using transit spectroscopy in the UV at Lyman-$\alpha$ \cite{vidal03}.  At
Lyman-$\alpha$ wavelengths, HD209458b is $\sim 3$ times larger than at optical
wavelengths.  The extended atmosphere may also contain carbon and oxygen as
reported by \cite{Madjar2004}.   Strong water absorption (Fig. \ref{tspec}) was
identified in the HD 209458b STIS data \cite{Barman2007} and later in emission
on the dayside with secondary eclipse Spitzer observations \cite{Knutson2008c}.
A marginal detection of water in HD189733 b was made with
Spitzer \cite{Tinetti2007} in the IR and later detected at much higher
confidence (along with methane) with NICMOS on board HST \cite{Swain2008b}.
Ground-based observations have also successfully repeated the Na detection in
HD209458 b \cite{Snellen2008}, and have actually resolved the doublet into its
two components for HD189733 b \cite{Redfield2008}.

Higher up in the atmosphere, at lower pressures and densities, global
circulation patterns become inefficient at redistributing heat and the
temperatures are likely close to their radiative equilibrium values.  Under
radiative equilibrium, there will be significant horizontal temperature
variations caused by different incident angles of incoming radiation from the
host star.  Furthermore, the photoionization of species like sodium and
potassium has been shown to be very important for modelling the transmitted
spectrum \cite{Fortney2003, Barman2007}.  Given the transparency of the
upper limb, photoionization (and potentially more complicated photochemistry)
may also spill over to the night side.  Consequently, {\em independent of the
global circulation}, there will be horizontal structure that is not likely well
reproduced in a 1D model of the transmitted spectrum. Efforts to model this
region in 3 dimensions are underway and should greatly improve our ability to
use transit spectroscopy to infer upper-atmospheric properties of
hot-Jupiters.

\section{\label{future}The future} 

The next decades promise a wealth of discoveries and new knowledge in the field of 
exoplanets.  Our understanding of exoplanet properties  will  progress
with expected developments on experimental and theoretical  fronts.
Ongoing and future high-pressure experiments in various national laboratories (Livermore and Sandia
in the US,  LIL and Laser Mega-Joule laser projects in France) and advancements of first principle
N-body numerical methods (DFT, path-integral, quantum molecular dynamics)
promise substantial progress on EOS in the critical pressure regime of 0.1-10 Mbar,
characteristic of H and He pressure ionization, and at
temperatures typical of planetary  interiors. The
 question of internal heat transport, and its efficiency in the presence of molecular weight gradients, may receive an answer from current progress in high-resolution multi-D numerical simulations.
 Numerical tools, such as the one developed by \cite{muthsam09} and devoted to high resolution
 stellar radiative hydrodynamical simulations, using grid refinement methods
 and high-order spatial schemes,
offer promising techniques to handle this problem. The future study of
exoplanet atmospheres is also poised for great progress in the coming years.
Coupled multi-D radiation hydrodynamical simulations are in the works, and new
and improved opacity databases are emerging.  Our understanding of clouds and
chemistry (much of which comes from studying brown dwarf atmospheres) is
steadily improving. 

From the observational standpoint,  a multitude of missions based on different techniques
and wavelengths will provide a wealth of data. The combination of all these techniques defines efficient strategies
for the detection and characterization of exoplanets, as highlighted by the impressive
prospective work by \cite{lunine08}. 
Current and future space-based surveys, like CoRoT and Kepler,
 will significantly increase
the number of know transiting systems. These two space missions may soon  answer to the question whether Earth analogs, in terms of mass and size, exist in the Habitable Zone. 
A large number of additional transiting planets will also  emerge in the next years
 from ongoing ground-based wide-field surveys as HATnet, TrES or WASP.
Increasing the number of detections will allow
a  more comprehensive study of exoplanet physical properties  (mass-radius relationship,
bulk composition, abnormal radii, etc..) and will confirm (or not) current observed trends, placing them
on firmer statistical footing.
 CoRoT is expected to find many tens of transiting giant planets and could discover
 a few tens (10-40) of super-Earths. CoRoT recently announced the detection of the very first transiting super-Earth candidate (still to be confirmed at the time of this writing), CoRoT-Exo-7b, with
 a radius of 1.7 $R_{\rm Earth}$ (\cite{leger09}). The Kepler mission, launched in  March 2009, will observe
hundreds of thousands of stars for transiting events and
 is expected to find about 1000 transiting gas giants. This mission will certainly provide the most reliable way to get a census of Earth-sized planets around solar type stars.
 It is capable of detecting several hundreds super-Earth in all orbits up to 1 year and more than 50 Earth-sized planets (1 $\rearth$) in all orbits, with about a dozen in $\sim$ 1-year period orbits (\cite{borucki03, valencia07, selsis07}). Kepler will thus be able to discover a large number of planets, but only a sub-sample of the most suitable ones 
 will be followed-up with Doppler velocimetry from the ground (HARPS-NEF and possible future instruments).  \cite{valencia07} estimate that Kepler, limited only by available observing time for HARPS-NEF, could obtain, for dozens of super-Earths, their radius and mass with a precision of
 less than 5\% and 10\% respectively. Such an accuracy allows bulk estimates of the composition
 for planets in the mass range 1-10 $\mearth$ \cite{valencia07}. More details on observational uncertainties for both CoRoT and Kepler can be found in \cite{selsis07}.

 Also, Kepler and CoRoT promise optical
phase curves of short-period planets, while a refurbished Hubble Space
Telescope will bring new transmission spectra (from the UV to near-IR) and
near-IR phase curves.  Future missions like JWST and SOFIA also promise new and
exciting exoplanet atmosphere observations.  As the successor of HST and Spitzer, JWST will open new avenues
in transiting exoplanet science with the characterization of intermediate and low mass exoplanets
($\le \mnep$). Currently, about ten hot Jupiters have been observed with Spitzer, allowing
the analysis of their emergent spectra, and the Warm mission should double the number of detection \cite{deming09}. Spitzer could in principle detect hot super-Earths in favorable cases,
and JWST will extend these detections to "warm super-Earths" ($\sim 300K$). The latter mission
is also capable of measuring the day-night temperature difference of warm Earth-like planets
orbiting M-dwarfs, just as  done by Spitzer for several hot Jupiters \cite{deming09}. JWST should be able to obtain light curves of primary and secondary eclipses of 1 $\rearth$ terrestrial planet around main sequence stars with a high precision, allowing basic characterisation of the transiting exoplanet and the possibility to search for unseen planets \cite{clampin09}.

From the ground, the recent discovery of the
triple planetary-mass system orbiting HR 8799
\cite{Marois2008} has clearly demonstrated the potential of
ground-based direct imaging campaigns.
The development of a new generation of adaptive-optics (AO) systems, such as 
VLT-SPHERE or the Gemini Planet Imager  (GPI) augurs well for direct imaging of planets orbiting solar type stars, enabling direct detection of hundreds of warm Jovian planets in
the next decade (\cite{macintosh09}). These systems, which are
precursors for the next generation of extremely large telescopes with apertures around 30m,
attempt to reach in a near future contrasts up to 10$^8$, allowing the detection of young Jupiters of less than 100 Myr around solar type stars (\cite{oppenheimer09}). 
In the next years, AO coronographs should thus be sensitive to a broad range of
self-luminous giant planets in the ranges 1-10 $\mjup$, 4-40 AU and 10-10 000 Myr (\cite{macintosh09}).
High contrast imaging systems on Extremely Large telescopes (ELT) should achieve contrasts allowing the characterisation of any self-luminous planet at high signal-to-noise ratio (SNR) and spectral resolution. Even mature planets in the inner part of solar systems ($< 2$ AU) should become detectable through reflected light (\cite{macintosh09}).

Better characterization of our own solar system planets is also crucial
to carry on, with for instance
the Juno mission to Jupiter which should measure atmospheric abundances, in particular water,  and accurately map its gravitational and magnetic fields.
 As stressed in this review, the understanding of exoplanet physical properties is strongly linked to the understanding of their formation process. On this front, our knowledge will greatly improve with 
the possibility to detect recently formed giant planets either directly, through their 
thermal or accretion luminosity using AO techniques as above mentioned,
or indirectly by imaging the structures ({\it e.g} gaps) they should produce in their disks with projects such as ALMA. 

Finally, the possibility of  identifying habitable planets,  with the detection of water vapor and signs of chemical disequilibrium in their atmospheres, is given by projects
such as DARWIN/TPF and JWST. This could be one of the most significant and stimulating achievements of Science, as it may tell us that we are not alone in the universe. 


\ack
The authors are grateful to their anonymous referee for valuable comments
and appreciated contribution to improve the manuscript.
The authors thank F. Selsis, B. Levrard, J. Leconte and C. Winisdoerffer for helpful discussions.
We are also grateful to F. Selsis and D. Saumon for providing some of the figures presented in this review. I. B and G. C thank the astronomy department of the University
of St Andrews, where part of this work was completed within the "Constellation" network, for their warm hospitality. This work was
supported by the Constellation european network MRTN-CT-2006-035890
and the french ANR "Magnetic Protostars and Planets" 
MAPP project. T. B acknowledges support form the NASA Origins of Solar Systems program along with
the Hubble and Spitzer theory programs.

\end{document}